\documentclass[pdflatex,sn-mathphys]{sn-jnl}% Math and Physical Sciences Reference Style
\jyear{2022}%

\theoremstyle{thmstyleone}%
\theoremstyle{thmstyletwo}%

\theoremstyle{thmstylethree}%

\begin{document}

\title[3D-RAPID]{Parallelized computational 3D video microscopy of freely moving organisms at multiple gigapixels per second}

\author*[1,2,6]{\fnm{Kevin C.} \sur{Zhou}}\email{kevinczhou@berkeley.edu}
\author[2]{\fnm{Mark} \sur{Harfouche}}
\author[3]{\fnm{Colin L.} \sur{Cooke}}
\author[2]{\fnm{Jaehee} \sur{Park}}
\author[1]{\fnm{Pavan C.} \sur{Konda}}
\author[1]{\fnm{Lucas} \sur{Kreiss}}
\author[1]{\fnm{Kanghyun} \sur{Kim}}
\author[1]{\fnm{Joakim} \sur{Jönsson}}
\author[2]{\fnm{Jed} \sur{Doman}}
\author[2]{\fnm{Paul} \sur{Reamey}}
\author[2]{\fnm{Veton} \sur{Saliu}}
\author[1,2]{\fnm{Clare B.} \sur{Cook}}
\author[2]{\fnm{Maxwell} \sur{Zheng}}
\author[2]{\fnm{Jack P.} \sur{Bechtel}}
\author[2]{\fnm{Aurélien} \sur{Bègue}}
\author[5]{\fnm{Matthew} \sur{McCarroll}}
\author[4]{\fnm{Jennifer} \sur{Bagwell}}
\author[2]{\fnm{Gregor} \sur{Horstmeyer}}
\author[4]{\fnm{Michel} \sur{Bagnat}}
\author*[1,2,3]{\fnm{Roarke} \sur{Horstmeyer}}\email{roarke.w.horstmeyer@duke.edu}

\affil[1]{Department of Biomedical Engineering, Duke University, Durham, NC 27708, USA}

\affil[2]{Ramona Optics Inc., 1000 W Main St., Durham, NC 27701, USA}

\affil[3]{Department of Electrical and Computer Engineering, Duke University, Durham, NC 27708, USA}

\affil[4]{Department of Cell Biology, Duke University, Durham, NC 27710, USA}

\affil[5]{Department of Pharmaceutical Chemistry, University of California, San Francisco, CA, USA}

\affil[6]{Current affiliation: Department of Electrical Engineering and Computer Sciences, University of California, Berkeley, CA, USA}


\abstract{
To study the behavior of freely moving model organisms such as zebrafish (\textit{Danio rerio}) and fruit flies (\textit{Drosophila}) across multiple spatial scales, it would be ideal to use a light microscope that can resolve 3D information over a wide field of view (FOV) at high speed and high spatial resolution. However, it is challenging to design an optical instrument to achieve all of these properties simultaneously. Existing techniques for large-FOV microscopic imaging and for 3D image measurement typically require many sequential image snapshots, thus compromising speed and throughput. Here, we present 3D-RAPID, a computational microscope based on a synchronized array of 54 cameras that can capture high-speed 3D topographic videos over a 135-cm\textsuperscript{2} area, achieving up to 230 frames per second at throughputs exceeding 5 gigapixels (GPs) per second. 3D-RAPID features a 3D reconstruction algorithm that, for each synchronized temporal snapshot, simultaneously fuses all 54 images seamlessly into a globally-consistent composite that includes a coregistered 3D height map. The self-supervised 3D reconstruction algorithm itself trains a spatiotemporally-compressed convolutional neural network (CNN) that maps raw photometric images to 3D topography, using stereo overlap redundancy and ray-propagation physics as the only supervision mechanism. As a result, our end-to-end 3D reconstruction algorithm is robust to generalization errors and scales to arbitrarily long videos from arbitrarily sized camera arrays. The scalable hardware and software design of 3D-RAPID addresses a longstanding problem in the field of behavioral imaging, enabling parallelized 3D observation of large collections of freely moving organisms at high spatiotemporal throughputs, which we demonstrate in ants (\textit{Pogonomyrmex barbatus}), fruit flies, and zebrafish larvae.}


\maketitle

%%%%%%%%%%%%%%%%%%%%%%%%%%  body  %%%%%%%%%%%%%%%%%%%%%%%%%%
\section{Introduction}
Quantifying the behavior and locomotion of freely-moving model organisms, such as the fruit fly (\textit{Drosophila}) and zebrafish (\textit{Danio rerio}), is essential in a wide variety of applications, including neuroscience \cite{bellen2010100,oliveira2013mind,kalueff2014zebrafish}, developmental biology \cite{dreosti2015development}, disease modeling \cite{pandey2011human,sakai2018zebrafish}, drug discovery \cite{macrae2015zebrafish, maitra2019using}, and toxicology \cite{hirsch2003behavioral,bambino2017zebrafish}. Particularly for high-throughput screening in these applications, it is desirable to monitor the behaviors of tens or hundreds of organisms simultaneously, thus requiring high-speed imaging over large fields of view (FOVs) at high spatial resolution, and ideally with the ability to observe behavior in 3D. Such an imaging system would allow researchers to bridge the gap between microscopic phenotypic expression and natural, multi-organism behavior that manifest across more macroscopic scales, such as shoaling \cite{wright2006repeated,harpaz2021precise}, courtship and aggression behaviors \cite{dankert2009automated,robie2017machine}, exploration \cite{dunn2016brain,johnson2020probabilistic}, and hunting \cite{bianco2011prey,patterson2013visually,muto2013prey,bolton2019elements,johnson2020probabilistic}.

Common approaches for behavioral recording utilize 2D wide-field microscopes with low-magnification optics to cover as large a FOV as possible. However, due to physical space-bandwidth product (SBP) limitations of conventional optics \cite{lohmann1989scaling,zheng2014fourier,park2021review}, standard imaging systems are forced to accept a tradeoff between image resolution and FOV (that is, can only record at low resolution when observing a large FOV).
Such systems are commonly used to track the location of large populations of organisms in high-content screening applications for toxicology and pharmacology \cite{rihel2010zebrafish,mccarroll2019zebrafish,mathias2012advances}, but cannot record key morphological features and behavioral signatures that require high-resolution capture.
Techniques that enhance SBP to facilitate high-resolution imaging over large areas, such as Fourier ptychography (FP) \cite{zheng2013wide,konda2020fourier,zheng2021concept} and mechanical sample translation \cite{kumar2020whole,borowsky2020digital}, often require multiple sequential measurements, which compromises imaging speed and throughput. Approaches that perform closed-loop mechanical tracking to record single organisms freely moving in 2D with scanning mirrors \cite{grover2016flyception} or moving cameras \cite{johnson2020probabilistic} are not scalable and thus cannot longitudinally observe multiple organisms simultaneously.

\begin{figure}[ht]
    \centering
    \centerline{\includegraphics[width=1.25\textwidth]{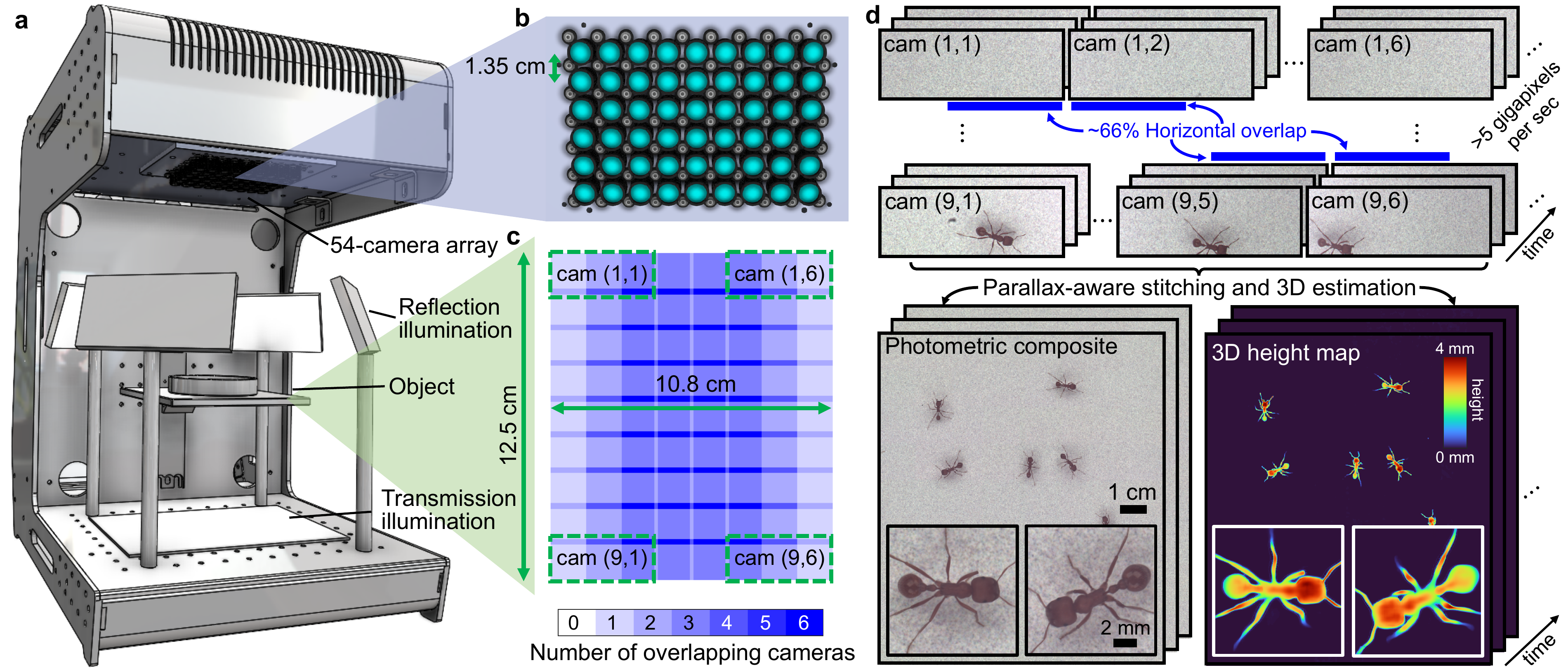}}
    \caption{Overview of 3D-RAPID. \textbf{a}, Computational microscope setup, consisting of a 9$\times$6 = 54 array of finite-conjugate imaging systems, jointly recording across a 135-cm\textsuperscript{2} area. LED arrays serve as the illumination source, both in transmission and reflection. \textbf{b}, 9$\times$6 array of cameras and lenses. \textbf{c}, Overlap map of the object plane, demonstrating roughly 66\% horizontal overlap redundancy between neighboring cameras (and minimal overlap in the vertical dimension). Four example camera FOVs are denoted with green dotted boxes, identified by (row,column) coordinates. \textbf{d}, The MCAM captures 54 synchronized videos at $>$5-GP/sec throughputs, which are stitched to form a high-speed video sequence of globally-consistent composites and the corresponding 3D height maps. }
    \label{fig:teaser}
\end{figure}

Conventional wide-field techniques also lack 3D information, which potentially precludes observation of important behaviors, such as vertical displacement and out-of-plane tilt changes in zebrafish larvae \cite{ehrlich2017control,ehrlich2019primal,bolton2019elements} and 3D limb coordination and kinematics in various insects \cite{akitake2015coordination, shamble2017walking, gunel2019deepfly3d, lobato2022neuromechfly}. Commonly used 3D microscopy techniques such as diffraction tomography \cite{wolf1969three,lauer2002new,horstmeyer2016diffraction,chowdhury2019high,zhou2020diffraction}, light sheet microscopy \cite{ahrens2013whole,chen2014lattice,patel2022high}, and optical coherence tomography (OCT) \cite{huang1991optical,zhou2019optical,zhou2021unified,zhou2022computational}, are not well-suited for behavioral imaging, since they often require multiple sequential measurements for 3D estimation and inertially-limited scanners that sacrifice speed. Furthermore, while such techniques can achieve micrometer-scale spatial resolutions, they typically do so over millimeter-scale FOVs rather than the multi-centimeter-scale FOVs necessary for imaging freely-moving organisms. Thus, these techniques are typically limited to imaging one immobilized organism at a time (e.g., embedded in agarose, tethered \cite{gunel2019deepfly3d,lobato2022neuromechfly}, or paralyzed), which prevents behavior studies.

Parallelized, camera array-based imaging systems have also been proposed to increase imaging system SBP and overall measurement throughput \cite{wilburn2005high,brady2012multiscale,lin2015camera,fan2019video,thomson2021gigapixel}; however, none of these prior approaches have demonstrated scalable, high-speed, high-resolution, wide-FOV, 3D imaging. In particular, several of these approaches were designed for 2D macroscopic photographic applications, which face several challenges for miniaturization for microscopy applications, or feature a primary objective lens that limits the maximum achievable system SBP (see Discussion). Various macroscale 3D depth imaging techniques have also been developed, such as time-of-flight light detection and ranging (LiDAR) \cite{jiang2020time}, coherent LiDAR \cite{riemensberger2020massively,okano2020swept,rogers2021universal, qian2022video, lukashchuk2022dual}, structured light \cite{geng2011structured}, stereo vision \cite{aguilar1996stereo}, and active stereo vision techniques \cite{scharstein2003high}. However, such 3D imaging systems have throughputs typically limited to 10s of megapixels (MPs) per second and generally have poor spatial resolutions on the order of millimeters, making them ill-suited for behavioral imaging of small model organisms. Further, active patterned illumination techniques do not scale to high pixel counts, typically require multiple measurements (thus compromising speed), and may directly impact the organism's behavior.

Here, we present \textbf{3D} \textbf{R}econstruction with an \textbf{A}rray-based \textbf{P}arallelized \textbf{I}maging \textbf{D}evice (3D-RAPID), a new computational 3D microscope based on an array of 9$\times$6 = 54 temporally synchronized cameras, capable of acquiring continuous high-speed video of dynamic 3D topographies over a 135-cm\textsuperscript{2} lateral FOV at 10s of micrometer 3D spatial resolution and at spatiotemporal data rates exceeding 5 gigapixels (GPs) per second (Fig. \ref{fig:teaser}). We demonstrate three operating modes of our microscope, which can be flexibly chosen depending on whether to prioritize speed (up to 230 frames per second (fps)) or spatial SBP (up to 146 MP/frame). We also present a new scalable computational 3D reconstruction algorithm that, for each synchronized snapshot, simultaneously forms a globally-consistent photometric composite and a coregistered 3D height map based on a ray-based physical model. The 3D reconstruction itself trains an underparameterized, spatiotemporally-compressed convolutional neural network (CNN) that maps multi-ocular inputs to the 3D topographies, using ray propagation physics and consistency in the overlapped regions as the only supervision. Thus, after computational reconstruction of just a few video frames ($<$20), 3D-RAPID can rapidly generate photometric composites and 3D height maps for the remaining video frames non-iteratively.

3D-RAPID thus solves a longstanding problem in the field of behavioral imaging of freely moving organisms that previously only admitted low-throughput solutions. To the best of our knowledge, prior to our work, there was no imaging system that could sustainably image at such high spatiotemporal throughputs ($>$5 GP/sec) in 3D. These new capabilities have allowed us to capture novel 3D measurements of freely moving organism behavior, which we have extensively tested in a series of experiments with three model organisms: zebrafish larvae, fruit flies, and ants. In particular, the large FOV of 3D-RAPID enabled imaging of multiple freely behaving organisms in parallel, while the dynamic 3D reconstructions and high spatial resolution and imaging speeds enabled 3D tracking of fine features, such as ant leg joints during exploration, zebrafish larva eye orientation during feeding, and fruit fly pose while grooming.

\section{High-throughput 3D video with 3D-RAPID}

\subsection{3D-RAPID hardware design}
The 3D-RAPID hardware is based on a multi-camera array microscope (MCAM) architecture \cite{harfouche2022multi, thomson2021gigapixel}, consisting of 54 synchronized micro-camera units spaced by 13.5 mm and tiled in a 9$\times$6 configuration. Each micro-camera captures up to 3120 $\times$ 4208 pixels (1.1-\textmu m pitch), for a total of $\sim$700 megapixels per snapshot. The data is transmitted to computer memory via PCIe at $\sim$5 GB/sec. 
Unlike conventional microscopy, 3D-RAPID is configured to acquire multi-view videos. That is, almost every point in the synthesized $\sim$12.5$\times$10.8-cm\textsuperscript{2} is viewed from at least two perspectives. To achieve this, we axially positioned the lenses (Supply Chain Optics, $f$ = 26.23 mm) to obtain a magnification of $M\approx0.11$, leading to $\sim$66\% overlap in the sample plane field of view (FOV) between cameras adjacent along the longer camera dimension (Fig. \ref{fig:teaser}c). This overlap redundancy enables 3D estimation using stereoscopic parallax cues.
The sample is illuminated in transmission or reflection using planar arrays of white LEDs covered by diffusers (Fig. \ref{fig:teaser}a).

\subsection{Tradeoff space of lateral resolution, field of view, and frame rate}
Our 3D-RAPID system has flexibility to downsample or crop the individual sensor pixels or use fewer cameras to increase the frame rate.
The overall data throughput is limited by the slower of two factors: the data transfer rate from the sensors to the computer RAM ($\sim$5 GB/sec) or the sensor readout rate, which is a function of the sensor crop shape and downsample factor. 
Streaming all 54 cameras without downsampling or cropping runs into the data transfer rate-limited frame rate of $\sim$7 fps. To achieve higher frame rates, we present results with a 1536$\times$4096 sensor crop using either 4$\times$, 2$\times$, or no downsampling, allowing us to achieve up to 230, 60, or 15 fps, respectively, while maintaining roughly the same overall throughput of $\sim$5 GP/sec (Table \ref{table:configurations}). While excluding half of the sensor rows all but eliminates FOV overlap in the vertical dimension, the benefits are two-fold: increased frame rate and reduced rolling shutter artifacts (see Methods~\ref{rolling_shutter}). 

\begin{table}
\centering
\begin{tabular}{ c | c c c}
 Downsample factor & 1$\times$ (none) & 2$\times$ & 4$\times$\\ 
 \hline
 Per-camera dims & 1536$\times$4096 & 768$\times$2048 & 384$\times$1024 \\  
 Composite dims & 13000$\times$11250 & 6500$\times$5625 & 3250$\times$2810\\
 Composite SBP & 146.3 MP & 36.6 MP & 9.1 MP\\
 Frame rate & 15 fps & 60 fps & 230 fps\\
 Exposure & 20 ms & 5 ms & 2.5 ms\\
 Raw pixel rate & 5.1 GP/sec & 5.1 GP/sec & 4.9 GP/sec\\
 Composite pixel rate & 2.2 GP/sec & 2.2 GP/sec & 2.1 GP/sec\\
 Image pixel pitch & 9.6 \textmu m & 19.2 \textmu m & 38.4 \textmu m \\
\end{tabular}
\caption{The three imaging configurations.}
\label{table:configurations}
\end{table}

\begin{figure}[ht]
    \centering
    \centerline{\includegraphics[width=1.1\textwidth]{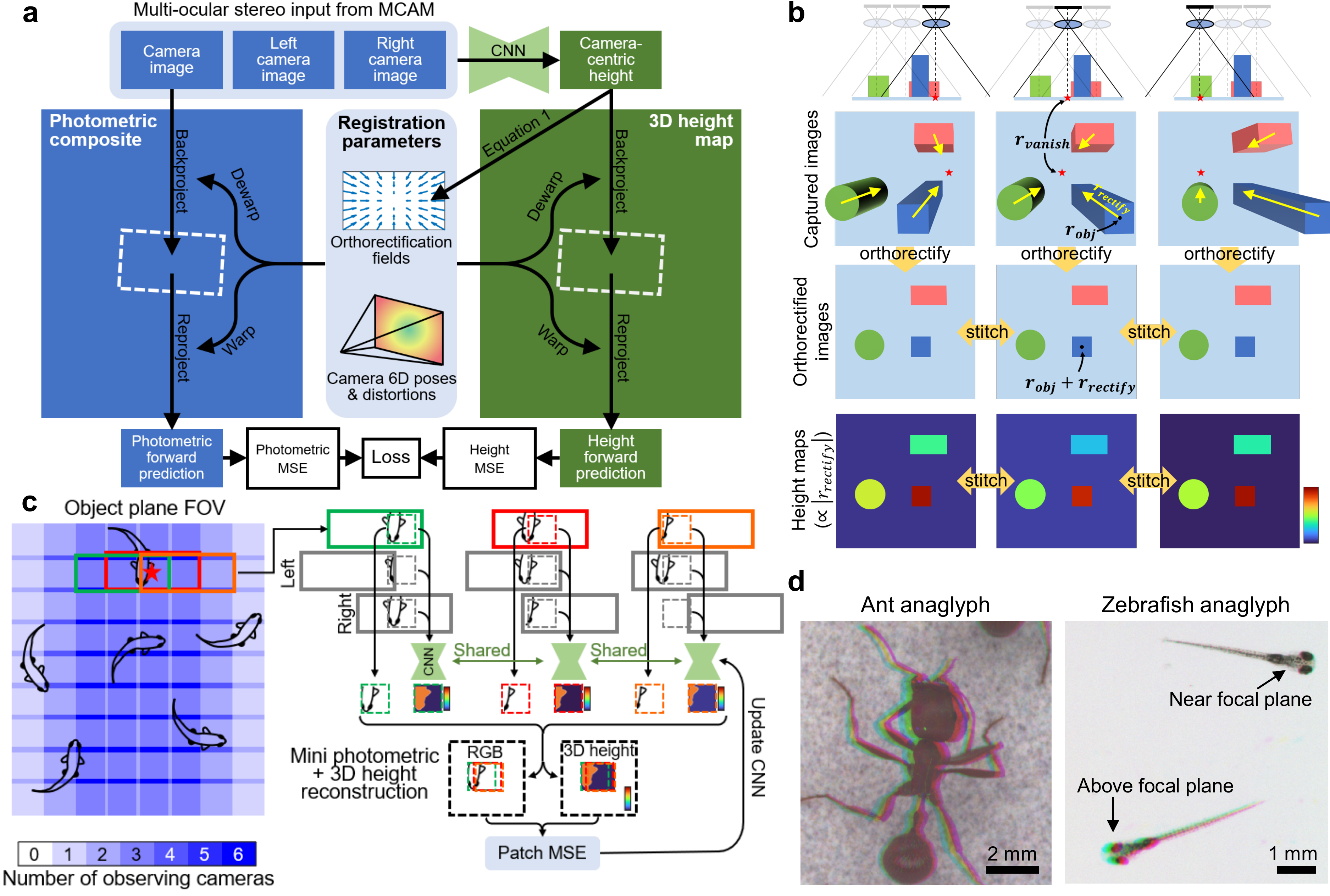}}
    \caption{Computational 3D reconstruction and stitching algorithm for 3D-RAPID. \textbf{a}, The algorithm starts with raw RGB images (only one shown for clarity), along with coregistered images from the cameras to left and right, as CNN inputs. CNN generates camera-centric height maps, which in turn dictate orthorectification fields (see \textbf{b} and Eq. \ref{orthorectification}). Orthorectificaton fields and camera poses + distortions constitute registration parameters, dictating where and how each image should be backprojected in the stitched photometric composite and 3D height map. The backprojection step is then reversed (reprojection) to form forward predictions of the RGB images and camera-centric height maps. Errors (photometric MSE and height MSE) guide the optimization of the CNN. \textbf{b}, The physical ray model, intuitively showing how orthorectification facilitates stitching of non-telecentric images and height maps. \textbf{c}, The patch-based joint training/stitching/3D reconstruction algorithm. At each gradient descent iteration, random coordinates are chosen (red star); all cameras that view a given point are isolated. A patch is cropped out from each camera image surrounding the randomly sampled point, along with the corresponding left/right camera images to serve as the multi-ocular stereo inputs to the CNN to predict the patch height map. These patches undergo the procedure outlined in \textbf{a} to form a mini photometric and 3D height reconstructions to update the CNN. Zeros are assigned to stereo input pixels when unavailable (e.g., at the edge of the object plane FOV), to preserve convolutionality when applying the CNN to the entire camera images to generate the full-size reconstructions. \textbf{d}, Analyphs, whereby the three stereo inputs are color-coded as RGB channels, showing the parallax that is used to estimate 3D.}
    \label{fig:methods}
\end{figure}

\subsection{Seamless image registration, stitching, and 3D estimation}\label{stitching}
For each video frame, the 3D-RAPID algorithm fuses the 54 synchronously acquired images, via gradient descent using a pixel-intensity-based loss, into a continuous, seamless, expanded-FOV composite image, and simultaneously estimates a coregistered 3D height map (Fig. \ref{fig:methods}a). In fact, these two tasks are intimately related -- to form a high-quality registration, it is necessary to account for parallax distortions induced by height deviations from a planar sample scene that would otherwise thwart simple registration using homographic transformations (Fig. \ref{fig:methods}b) \cite{kumar1994direct,sawhney19943d,zhou2021mesoscopic}. 
To achieve this, the algorithm starts with calibration of the 6-degree-of-freedom poses ($x$, $y$, $z$, roll, pitch, yaw), camera distortions, and intensity variations by registering and stitching 54 images of a flat, patterned target (Methods \ref{calibration}). 
Estimating the 3D height map of the sample of interest relative to this calibration plane is tantamount to rendering the images registerable using homographies (Fig. \ref{fig:methods}b). In particular, the per-pixel deformation vectors that undo the parallax shifts (i.e., \textit{orthorectify} the images) have magnitudes that are directly proportional to the per-pixel heights, $h(\mathbf{r})$ (i.e., the height map), given by \cite{zhou2021mesoscopic}
\begin{equation}\label{orthorectification}
    h(\mathbf{r_\mathit{obj}}+\mathbf{r_\mathit{rectify}})=f\frac{\|\mathbf{r_\mathit{rectify}}\|}{\|\mathbf{r_\mathit{obj}}-\mathbf{r_\mathit{vanish}}\|} \left(1+\frac{1}{M}\right)
\end{equation}
where $f=26.23$ mm is the effective focal length of the lens, $M\approx0.11$ is the linear magnification, $\mathbf{r_\mathit{obj}}$ is the apparent 2D position of the object in the pixel (before orthorectification), $\mathbf{r_\mathit{vanish}}$ is the vanishing point to which all lines perpendicular to the sample plane appear to converge, and $\mathbf{r_\mathit{rectify}}$ is the 2D orthorectification vector pointing towards the vanishing point (Fig. \ref{fig:methods}b). $\mathbf{r_\mathit{vanish}}$ can be determined from the camera pose, as the point in the sample plane that intersects with the perpendicular line that passes through the principal point in the thin lens model.
The orthorectification vectors $\mathbf{r_\mathit{rectify}}$, and therefore the height map, for each object position $\mathbf{r_\mathit{obj}}$ can be determined by registering images (via photometric pixel values) from different perspectives. The accuracy of the height map thus depends on the object having photometrically textured (i.e., not uniform) surfaces that enable unique image registration, a condition which the model organisms we imaged satisfied.

Thus, the optimization problem is to jointly register all 54 images using the pixel-wise photometric loss, using the orthorectification maps (which are directly proportional to the height maps via Eq. \ref{orthorectification}) as the deformation model on top of the fixed, pre-calibrated camera parameters, including distortions (Fig. \ref{fig:methods}a,b). In practice, since viewpoint-dependent photometric appearance can affect image registration, we also employed normalized high-pass filtering to standardize photometric appearance (Methods \ref{illumination_robust} and Supplementary Sec. \ref{hpf}).

\subsection{Spatiotemporally-compressed 3D video via end-to-end physics-supervised learning}
\label{DL}

Instead of optimizing the height maps directly,
we reparameterized the height maps as the output of a fully-convolutional encoder-decoder CNN that takes the multi-view stereo images as inputs.
This reparameterization has two interpretations, depending on whether we emphasize the CNN or the ray-based physical model. On the one hand, the CNN can be thought to act entirely as a training-data-free regularizer (i.e., deep image prior (DIP) \cite{ulyanov2018deep}) that safeguards against 3D reconstruction artifacts that may otherwise arise from practical deviations from modeling assumptions that thwart image registration \cite{zhou2021mesoscopic}. For example, using the CNN as a regularizer can be useful when the sample has a different appearance when viewed from different angles, which can be caused by uneven illumination, angle-dependent scattering responses, or varying pixel responses.
Since we wish to reconstruct hundreds to thousands of 3D video frames, it would be prohibitively slow to independently reconstruct every individual video frame, with or without CNNs. Thus, we use one shared DIP, with each frame encoded by the raw multi-ocular stereo photometric inputs. 

On the other hand, this leads to the second interpretation of a self-supervised or physics-supervised learning problem, in which the image registration of the overlapped MCAM image frames, governed by a ray-based thin lens physical model (Eq. \ref{orthorectification}), provides the physics-based supervision that guides the CNN training (Fig. \ref{fig:methods}a,c). The CNN can then be used to generalize to other MCAM data, both spatially (other micro-cameras) and temporally (other video frames).

This dual interpretation of our CNN-regularized, physics-supervised learning approach reveals several advantages.
First, since we employ a fully-convolutional CNN, we can optimize on arbitrarily-sized image patches (Fig. \ref{fig:methods}c) that can fit in GPU memory, and then perform non-iterative forward inference on arbitrarily-large full-size images (Fig. \ref{fig:cnn_inference}). Thus, our proposed approach is scalable and generalizable to arbitrarily many cameras, each with arbitrarily many pixels, for arbitrarily many video frames. For implementation details on patch-based training, see Sec. \ref{patch}, Fig. \ref{fig:methods}c, and Supplementary Sec. \ref{patch_detailed}. Second, the CNN acts as a spatiotemporally-compressed representation of the 3D height map videos, thus avoiding the need to iteratively optimize every single 3D video frame.
Third, this spatiotemporal compression offers additional regularizing effects on top of the dataset-free, DIP-based regularization. As there are far fewer parameters in the CNN than height map pixels across all MCAM video frames, overfitting becomes less likely. Furthermore, the CNN implicitly enforces consistency across space and time, thus, for example, avoiding variance induced by independent optimization runs on different frames. Fourth, our approach has an inherent fail-safe against CNN generalization errors, unlike other deep learning-based approaches, since the ground truth is implicitly always available via the overlap redundancy of the MCAM along with the physical model.

\subsection{Patch-based learning from multi-ocular stereo inputs} \label{patch}

While Fig. \ref{fig:methods}a summarizes the ideal joint 3D reconstruction, stitching, and training method, in practice we are constrained by GPU memory. Thus, we train the CNN using a random patch sampling approach (Fig. \ref{fig:methods}c). Briefly, at each optimization iteration, we sample $n_\mathit{batch}$ (batch size) random points within the composite FOV (one shown in Fig. \ref{fig:methods}c). All cameras viewing each point are selected, from which patches surrounding that point are extracted from each camera view. Thereafter, these $n_\mathit{batch}$ groups of selected patches independently undergo the procedure outlined in Fig. \ref{fig:methods}a. 
Once CNN training is done, the backprojection step in Fig. \ref{fig:methods}a is carried out for each full temporal frame to create the stitched RGBH 3D reconstructions (Fig. \ref{fig:cnn_inference}). For more implementation details, see Supplementary Sec. \ref{patch_detailed}.

As mentioned in the previous section (Sec. \ref{DL}), the CNN is supplied multi-view inputs of the same sample scene (as shown in Fig. \ref{fig:methods}a,c), whose goal is to improve the generalizability of the CNN. These neighboring views are stacked along the channel input dimension in a way that preserves convolutionality, so that patch training and full-FOV inference are consistent (Supplementary Sec. \ref{patch_detailed}). This is beneficial because monocular stereo depth estimation is insufficient for objects whose appearances don't change significantly as a function of depth. For example, when imaging a fruit fly or zebrafish larva, it is difficult to distinguish between height-dependent magnification changes and natural variation in organism size. Thus, we train our CNN to solve a multi-ocular stereo 3D estimation problem, which is better-posed, as the 3D supervision signal itself is derived from the registration of the multi-ocular data (Supplementary Sec. \ref{monocular}). In this paper, we use 3 stereo inputs or fewer (center, left, and right, if available).

\begin{figure}[ht]
    \centering
    \centerline{\includegraphics[width=1.2\textwidth]{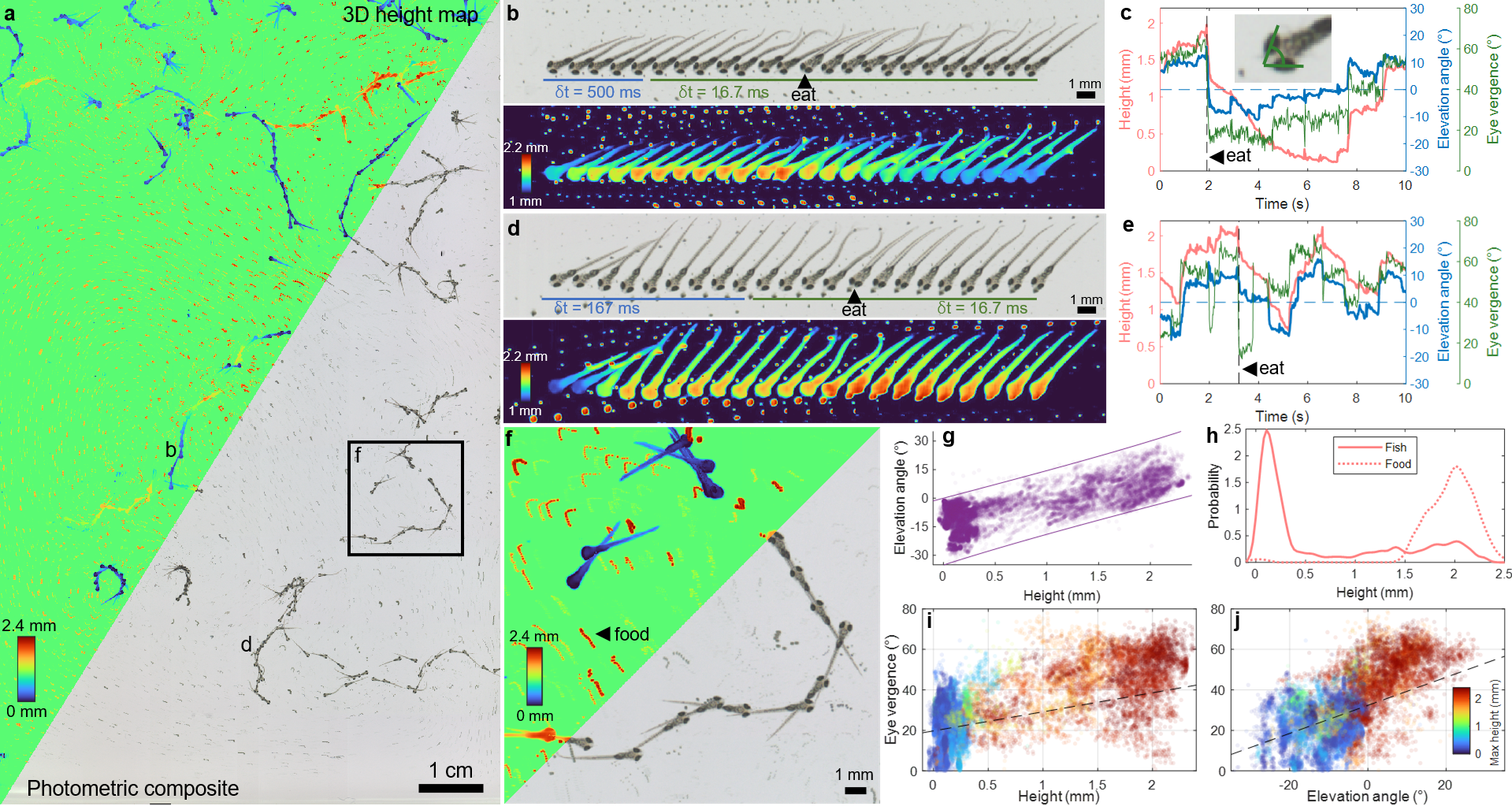}}
    \caption{Zebrafish larvae (10 dpf) swimming in an open arena with interspersed microcapsulated food particles (AP100), acquired at 60 fps for 10 sec (Supplementary Videos 1-3). \textbf{a}, 3D height map and photometric composites of the zoomed-out FOV, projected across every 50th temporal frame (0.83 sec) to highlight dynamics. The height map assigns an arbitrary value to the otherwise empty background. \textbf{b}, Photometric and height map frames of a single tracked fish feeding on AP100. The first 5 frames are spaced by 500 ms while the remaining frames are spaced by 16.7 ms (the full frame rate). \textbf{c}, The same fish's head height, elevation angle (pitch), and eye vergence angle (illustrated in inset) throughout the 10-sec video. \textbf{d}-\textbf{e}, Another example of a zebrafish feeding event. Note the change in eye vergence before and after the feeding event in both \textbf{b} and \textbf{d}. \textbf{f}, A zoomed-in region of \textbf{a}, showing 3 individual larvae in varying states of activity. The small red tracks are the drifting and floating AP100 food particles. \textbf{g} Fish head height vs. elevation angle for all 40 fish over time. Lines define the approximate physical limits due to geometric fish mobility constraints. \textbf{h}, Kernel density estimates of the height distributions of the zebrafish and AP100 food particles. Eye vergence vs. head height (\textbf{i}) and vs. elevation angle (\textbf{j}) plots are color-coded by the maximum height the fish attained in the 10-sec video. Fixed effect components of the linear mixed-effects regression lines are plotted ($p = 0.33$ and $p<10^{-5})$ for \textbf{i} and \textbf{j}, respectively).
    }
    \label{fig:zebrafish}
\end{figure}

\section{Results}

\subsection{3D-RAPID system characterization}
Our 3D-RAPID system has a full-pitch lateral resolution of $\sim$25 \textmu m and DOF of $\sim$9.4 mm, based on imaging a USAF resolution target and translating a patterned target axially (see Supplementary Sec. \ref{supp_characterization}). We validated the height precision and accuracy of our 3D-RAPID system by imaging precisely machined (to within 0.3 \textmu m) and interferometrically characterized gauge blocks (Mitutoyo). As expected, accuracy and precision of the reconstructed height improve when imaging at higher spatial resolution, which facilitates more accurate measurement of parallax shifts (see Supplementary Sec. \ref{hardware_theory}). Specifically, we achieved sub-20 \textmu m accuracy and precision in the 15-fps configuration, and $\sim$37 \textmu m and $\sim$74 \textmu m accuracy and precision in the 230-fps configuration. See Supplementary Sec. \ref{supp_characterization} for detailed characterization results.

\subsection{Zebrafish larvae (\textit{Danio rerio})}
We applied 3D-RAPID to several 10-sec videos of zebrafish larvae (\textit{Danio rerio}) freely swimming in a large 97 mm $\times$ 130 mm open arena using the 60-fps and 230-fps configurations (Table \ref{table:configurations}) across three separate experiments, the first of which was on 10-dpf fish feeding on microcapsule food particles (AP100) (Supplementary Videos 1 (60 fps), 2 (230 fps), and 3 (60 fps with tracking)). 
Fig. \ref{fig:zebrafish} summarizes the results for the 60-fps video of the 10-dpf fish feeding on AP100, most of which are floating at or near the water surface (Fig. \ref{fig:zebrafish}h). We tracked all 40 fish using a simple particle-tracking algorithm (Methods \ref{tracking}; Supplementary Video 3). The high throughput of 3D-RAPID allowed us to observe fine detail over a very wide FOV, capturing multiple rapid feeding events ($\sim$10s of ms), as shown in Fig. \ref{fig:zebrafish}b,d. From the photometric images, we can see that the larvae turn their bodies laterally so that their ventrally positioned mouths can access the overhead floating food. 
We also observe eye convergence once the larvae identify and approach the target, as shown in previous studies \cite{bianco2011prey, patterson2013visually, muto2013prey}. The eye angles rapidly deconverge after food capture (Fig. \ref{fig:zebrafish}c,e). The older fish (20 dpf) exhibit similar eye behavior when feeding on brine shrimp (Supplementary Videos 4, 5).

The 3D topographic information enabled by our technique reveals how the larvae axially approach their targets from below, including their head heights and elevation (pitch) angles during these feeding events (Fig. \ref{fig:zebrafish}b-e) \cite{bolton2019elements}. Note that the larvae's head height matches that of the targeted food particle during ingestion (see also in Supplementary Videos 1, 2, 4, 5), offering validation of our technique.

In addition to making organism-level observations, the high throughput of 3D-RAPID enabled us to make population-level inferences
by aggregating height and elevation angle information for all 40 individually-tracked larvae for all in-frame time points. The results show a roughly linear trend between height and elevation angle (Fig. \ref{fig:zebrafish}g), which can be explained based on the mobility constraints defined by the length of the larvae and the water depth. For example, if the head is at the bottom of the arena, then the elevation angle must be negative. Assuming a larval length of $L=4$ mm and a water depth of $H=2.3$ mm, these geometric constraints on the elevation angle, $\phi$, for a fish at height, $h$, are
\begin{equation}
    \phi_\mathit{min}(h) = \sin^{-1}(h/L), \ \ \ \ 
    \phi_\mathit{max}(h) = \sin^{-1}((H-h)/L),
\end{equation}
which are plotted in Fig. \ref{fig:zebrafish}g. This offers additional validation of the accuracy of our 3D height maps, suggesting future applications in studying fish locomotion dynamics \cite{ehrlich2019primal}. We also estimated the probability distributions of the heights of the larvae and the food particles (Fig. \ref{fig:zebrafish}h), both of which are bimodal. Predominantly, the larvae dwell at the bottom of the arena, only occasionally venturing upwards to hunt or forage for food.

Finally, we also analyzed population-level correlations between eye vergence angle (Methods \ref{tracking}), a property observable in the photometric images, and the fish height and elevation angle, which are derived from our 3D height maps (Fig. \ref{fig:zebrafish}i,j), across $n=39$ fish (one stationary fish excluded). Specifically, we used a linear mixed-effects model, where height or elevation angle is the fixed effect and dependence among images from the same fish are accounted for as random effects. Analyses of variance suggest that while fish height is not a statistically significant linear predictor of eye vergence angle ($p=0.33$), fish elevation angle is ($p<10^{-5}$). This is consistent with the fact that 
when the fish is swimming upwards, it is likely focusing on a food particle close to the surface. On the other hand, the fish can still be close to the surface following a feeding event, immediately after which the eyes deconverge (Fig. \ref{fig:zebrafish}b-e).

With the 230-fps configuration of our system, we can trade off spatial resolution to temporally resolve higher-speed zebrafish larval locomotion.
For example, compare the beginning of Supplementary Videos 6 and 7, which feature rapidly swimming zebrafish larvae, captured at 60 fps and 230 fps, respectively. Similarly, we can resolve the 4D fish dynamics as it attempts to swallow a live brine shrimp (Supplementary Videos 4 (60 fps) and 5 (230 fps)). 

\begin{figure}[t]
    \centering
    \centerline{\includegraphics[width=1.2\textwidth]{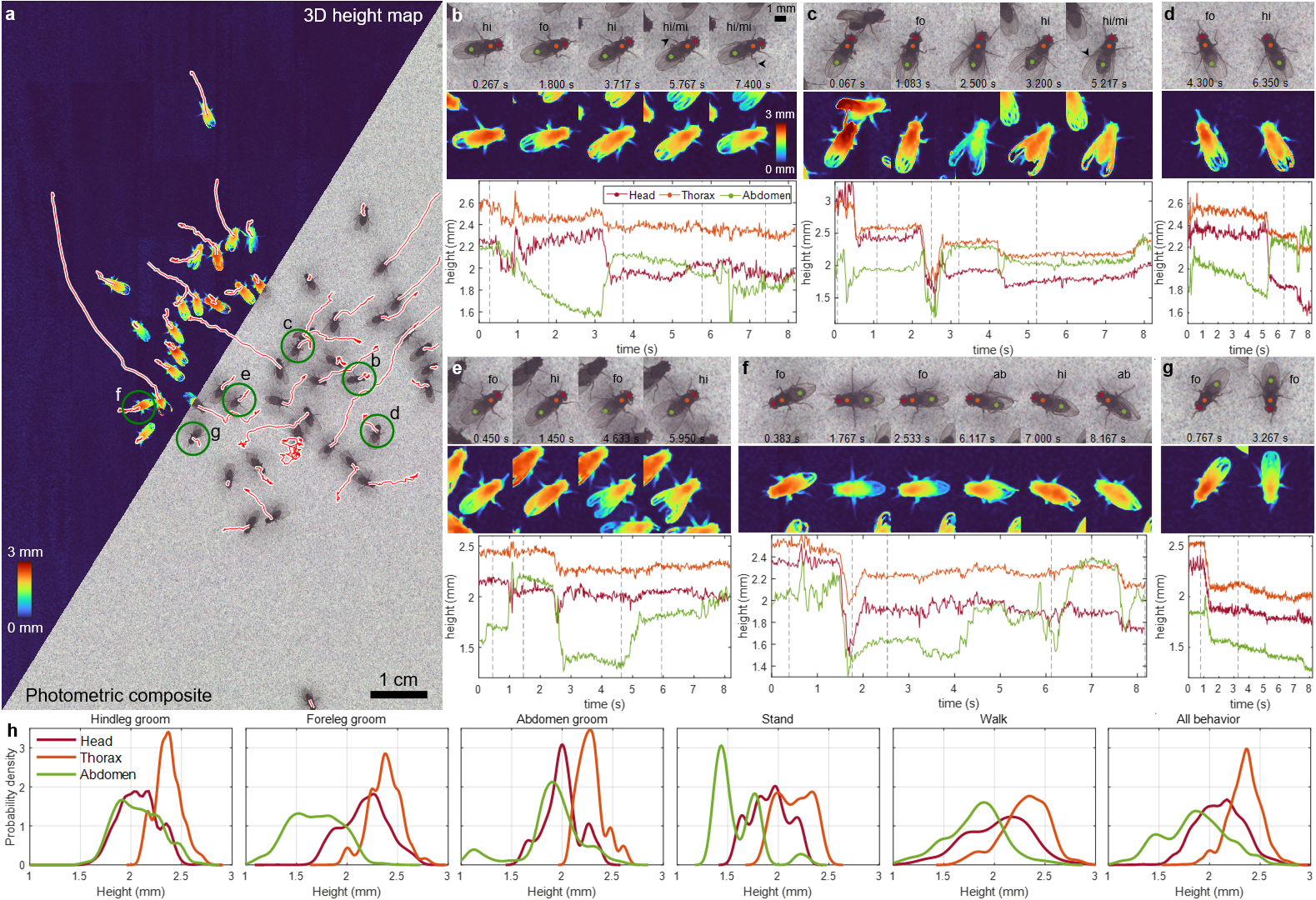}}
    \caption{Adult fruit flies freely moving across a flat, noise-patterned surface, acquired at 60 fps for 8 sec (Supplementary Videos 8-10). \textbf{a}, 3D height map and photometric composites of the zoomed-out FOV. The white-outlined red lines are the trajectories the 50 flies take. The green-circled flies are analyzed in the other figure panels. \textbf{b}, Select photometric and height map frames of a single tracked fly, exhibiting several grooming behaviors (hi = hindleg grooming, fo = foreleg or head grooming, mi = mid leg participation, ab = abdominal grooming). The time points of the frames are indicated by dotted lines in the plot below, which in turn highlights the changing heights of the head, thorax, and abdomen for the different grooming actions. \textbf{c}-\textbf{g}, The same information for 5 additional flies. \textbf{h} Kernel densities of the heights of head, thorax, and abdomen for various behaviors. Differences of head ($p<10^{-7}$), thorax ($p<10^{-16}$), and abdomen ($p<10^{-62}$) heights across behaviors are statistically significant (n = 43 flies).}
    \label{fig:flies}
\end{figure}

\subsection{Fruit flies (\textit{Drosophila hydei})}

Next, we applied 3D-RAPID to image and track 50 freely exploring adult fruit flies (\textit{Drosophila hydei}) under the 60-fps (Supplementary Videos 8 and 10) and 230-fps (Supplementary Video 9) configurations. Fig. \ref{fig:flies} summarizes the results for the 60-fps configuration for six individual flies exhibiting various behaviors. Supplementary Video 10 shows tracking of all 50 flies. The 3D height map offers additional insights into such grooming behaviors, building upon works that study freely-moving flies in 2D \cite{branson2009high, berman2014mapping} and 3D in single tethered flies \cite{gunel2019deepfly3d}. In particular, we observed changes in fly height and body tilt as the flies transition between different grooming behaviors. In Fig. \ref{fig:flies}b, as an individual fly transitions between grooming with its hindlegs and forelegs, the abdomen moves up and down, respectively, relative to the head and thorax. When a middle leg joins the grooming (Fig. \ref{fig:flies}b, arrowheads), there is a subtle change in abdomen height relative to head height. In Fig. \ref{fig:flies}c, our method correctly predicts an elevated height as one fly climbs atop another. At 2.5 sec, the fly's height drops, consistent with the straightened leg joints. A similar body tilt trend is observed for foreleg vs. hindleg grooming in this fly, as well as in Fig. \ref{fig:flies}d, e, and f. In Fig. \ref{fig:flies}f, we see another instance of the fly's leg joints fully extended at 1.767 sec, resulting in a reduced overall height. Further, we observe that the abdomen takes on a different relative height during abdominal grooming compared to hindleg grooming. Finally, in Fig. \ref{fig:flies}g, although the fly is grooming its forelegs throughout the video, it reduces its overall height after 1 sec, consistent with its extended leg posture.

To analyze population trends, we annotated video frames across $n=43$ flies flies with one of five behaviors: hindleg grooming, foreleg/head grooming, abdomen grooming, standing still, and walking (Fig. \ref{fig:flies}h). Flies that exited the FOV were excluded. We tested for cross-behavioral differences in heights of the head, thorax, and abdomen using three separate linear categorical mixed-effects models, accounting for random effects due to correlations among video frames from the same fly. Analyses of variance suggest that behavior groups are a statistically significant predictor of the heights of the head ($p<10^{-7}$), thorax ($p<10^{-16}$), and abdomen ($p<10^{-62}$).

\begin{figure}[ht]
    \centering
    \centerline{\includegraphics[width=1.2\textwidth]{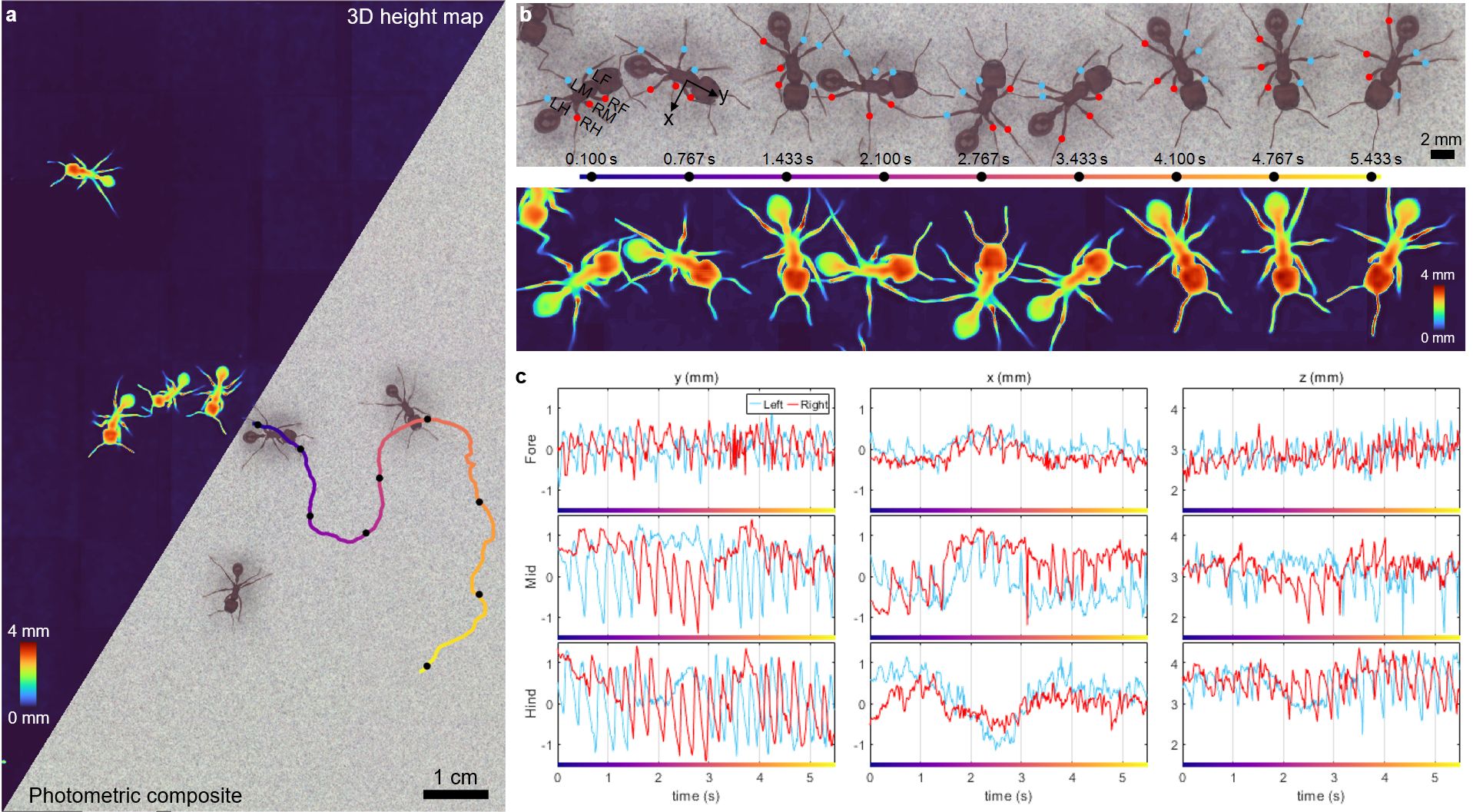}}
    \caption{Harvester ants freely moving across a flat, noise-patterned surface, acquired at 60 fps for 10 sec (Supplementary Videos 11-12). \textbf{a}, Photometric composite and 3D height map of the zoomed-out FOV. One of the ants' trajectories is color-coded by time, progressing from blue to red over a 5.5-sec duration, and is analyzed in \textbf{b} and \textbf{c}. \textbf{b}, Temporal snapshots of a single tracked ant along the trajectory in \textbf{a}. The blue and red dots are the femur-tibia joints for the ant's 6 legs (L = left, R = right, F = foreleg, M = middle leg, H = hindleg). \textbf{c}, The 3D positions of the femur-tibia joints over the 5.5-sec trajectory. The lateral dimensions ($xy$) are defined relative to the ant's orientation, as illustrated in \textbf{b}. }
    \label{fig:ants}
\end{figure}

\subsection{Harvester ants (\textit{Pogonomyrmex barbatus})}

We also imaged freely exploring red harvester ants (\textit{Pogonomyrmex barbatus}) under the 60-fps (Supplementary Video 11) and 230-fps (Supplementary Video 12) configurations. The 60-fps results are summarized in Fig. \ref{fig:ants}. From the static 3D height map frame, it is immediately obvious that the body is sloped downward, from the head to the abdomen \cite{reinhardt2014level}.
We used the dynamic 3D reconstructions enabled by 3D-RAPID to track the femur-tibia joints of all six legs of an individual ant (Fig. \ref{fig:ants}b,c; Methods \ref{tracking}), providing information about the kinematics of ant locomotion. The joint trajectories are plotted in Fig. \ref{fig:ants}c, showing that the high-frequency ($\sim$3-4 Hz) oscillations from walking kinematics are anti-correlated (180$^\circ$ out of phase) between left and right legs. This oscillation frequency remains relatively constant throughout the ant's journey. Further, the forelegs and hindlegs on the same side of the body are correlated, but anti-correlated with the mid legs on the same side of the body. These behaviors are consistent with the well-known alternating tripod gait pattern in ants \cite{zollikofer1994stepping,reinhardt2014level,shamble2017walking}, which persists even as the curvature of ant trajectory changes in our tracked ant. 

We also observe changes in lower-frequency gait patterns as the ant makes multiple turns throughout its exploration. In the first $\sim$1.5 sec, as the ant is turning right, we see a reduced oscillation amplitude in the mid and hindlegs on the right side in both the $y$ and $z$ directions; however, for the $x$ direction, we see the opposite trend (see Fig. \ref{fig:ants}b for the ant-centric coordinate system definition). Between 1.5 and 3 sec, as the ant is turning \textit{left}, we see the opposite motions as in the first 1.5 sec -- the oscillation amplitudes in the mid and hindlegs on the \textit{left} are reduced in both the $y$ and $z$ directions, while amplitude of the right mid leg motion in the $x$ direction is reduced. From 3 to 4.5 sec, the ant once again is turning right and we see similar trends as in the first 1.5 sec. Overall, this reduction in motion in $y$ and $z$ on the side of the ant corresponding to the direction the ant is turning is consistent with prior knowledge \cite{zollikofer1994stepping}. Interestingly, the amplitudes of the foreleg oscillations on both the left and right sides in both $y$ and $z$ remain relatively constant throughout the entire 5.5 sec, suggesting a lesser role in the biomechanics of changing directions. 

Finally, we observe a low-frequency oscillation (with a period of $\sim$4 sec) in the $x$ direction for all 6 legs that is correlated with the curvature of the ant's trajectory. Unlike the high-frequency (3-4 Hz) walking kinematics, which are anti-correlated between left and right, these low-frequencies are \textit{correlated} between left and right legs, suggesting left-right coordination when the ant is turning. These low-frequencies in the $x$ direction further are correlated between the forelegs and mid legs, but anti-correlated with the hindlegs.

\section{Discussion}

We have presented 3D-RAPID, a new computational microscope with a unique capability of dynamic topographic 3D imaging at 10s-of-\textmu m resolution and accuracy, over $>$130-cm\textsuperscript{2} FOV at throughputs exceeding 5 GP/sec. To handle the large data load, we devised an efficient, end-to-end, physics-supervised, CNN-based, joint 3D reconstruction and stitching algorithm that scales to arbitrarily long videos and arbitrarily sized camera arrays.
The high throughput of 3D-RAPID enabled us to study several freely-behaving model organisms at high speed and high resolution over a very large FOV. Thus, our technique fills a unique niche, enabling new ways for scientists to study small features of individual organisms over a large FOV that allows unconstrained social interactions of multiple organisms in parallel in 3D at high speed. For example, 3D-RAPID could be applied to study dueling behavior in ants \cite{yan2022insulin}, sexual behavior in fruit flies \cite{pavlou2016neural}, and feeding decisions in fruit flies \cite{sareen2021neuronal}.

3D-RAPID differs from other camera array-based techniques \cite{wilburn2005high,brady2012multiscale,lin2015camera,fan2019video,thomson2021gigapixel} in several ways, stemming from the challenge of adapting to microscopy applications. In particular, due to the large magnification requirements, the cameras need to be physically packed more tightly, which is a practical challenge due to mechanical constraints and heat dissipation management. Some approaches alleviate this challenge by using a primary objective lens to magnify the object to an intermediate image plane, which is then imaged by a camera array. However, this strategy limits scalability, as the primary objective’s intrinsic SBP would limit the total imaging throughput. Instead, we solved this problem by tiling all of the array’s CMOS sensors at the chip-scale onto a common multi-layer PCB, which is connected to a single FPGA for unified data routing. This allows for extremely tight packing and scalability by simply adding more sensors. Finally, 3D-RAPID also differs from light field imaging, because our cameras exhibit almost the theoretical minimum amount of overlap necessary for 3D surface estimation – this is an important design consideration because it allows us to maximize the SBP. In particular, to our knowledge, 3D-RAPID is the 3D imaging system with the highest sustained throughput to date. 

While we have presented several convincing 3D behavioral imaging demonstrations, there are several avenues for improvement. The hardware configuration could be adjusted to improve the 3D height reconstruction accuracy, which depends on how accurately parallax shifts can be detected to match corresponding features from different cameras.
In Supplementary Sec. \ref{hardware_theory}, we derive several equations detailing how height accuracy is impacted by hardware design parameters, suggesting that decreasing the focal length and increasing the magnification and sensor-to-sensor spacing improve height accuracy.
Furthermore, since the reconstruction algorithm is agnostic to the contrast mechanism, it would also be possible to incorporate other optical contrast mechanisms into 3D-RAPID, such as fluorescence to correlate behaviors with molecular signatures. Finally, throughput could be improved beyond 5 GP/sec by alleviating data transfer bottlenecks to the computer.

In summary, we have presented a high-throughput computational 3D topographic microscope as a new platform for studying the behavior of multiple freely-moving organisms at high speed and resolution over a very wide area. We expect our technique to be broadly applicable to elucidate new behavioral phenomena, not only in zebrafish, fruit flies, and ants, but also other model organisms such as tadpoles (\textit{X. laevis} and \textit{X. tropicalis}) and nematodes (\textit{C. elegans)}.

\section{Methods}

\subsection{Temporal synchronization of the camera array}
\label{rolling_shutter}
Ideally, all sensor pixels should be fully synchronized with a global shutter, not only within each sensor, but also across sensors. This would ensure that between different views of the same object, after accounting for camera poses, the only discrepancies are due to parallax shifts and not sample motion. For example, if two camera views of a moving object with zero height were desynchronized, lateral motion could be interpreted as a parallax shift, leading to an erroneous height estimate.
In practice, each of our sensors exhibits a rolling shutter, whereby only a single pixel value can be read out at a given time for a given sensor, row by row from the top-left to bottom-right corner in a raster scan pattern. This means that the bottom of a given sensor is captured later than the top of the sensor immediately below. However, across independent sensors, this rolling shutter readout pattern is synchronized to within 10 \textmu s, limited by the serial communication interface (I2C with a 100-kHz clock).

To mitigate the rolling shutter effects, we employed two strategies. First, we cropped the sensors so that there is only significant overlap in the horizontal dimension for stitching, in which the desynchronization is much less severe. Second, we calculated that with exposures of 2.5 ms for 4$\times$ downsampling, 5 ms for 2$\times$ downsampling, and 20 ms for no downsampling, artifacts would be minimal. For a detailed discussion and calculations, see Supplementary Sec. \ref{rolling_shutter_supp}.

\subsection{Achieving robustness to illumination variation}
\label{illumination_robust}
Since the optimization metric of our approach is the mean square per-pixel photometric error, we would achieve optimal performance when the sample has a camera-independent photometric appearance. This condition would require not only uniform response across all pixels of all cameras, but also that the sample is isotropically emanating light in all directions. The latter property is in practice difficult to achieve, requiring either perfectly diffuse illumination or a diffusely scattering sample, regardless of the illumination direction. In addition to the regularizing effects of the CNN/DIP, we employed two additional strategies to reduce the effects of camera-dependent appearance. First, as part of the camera pose pre-calibration procedure, we also jointly optimized per-camera second-order 2D polynomials (with cross terms) to correct the slowly-varying image intensity variation (whether caused by uneven illumination or camera response), using the same photometric stitching loss. Thus, the pre-calibration step not only ensures geometric consistency of the 54 images, but also photometric continuity. For more details, see Methods \ref{calibration}, below.

Second, for terrestrial organisms illuminated in reflection, we employed a two-step optimization process, where we first optimize the CNN to register the images using the RGB intensities. In the second step, we continue optimizing the CNN, except this time registering normalized high-pass-filtered versions of the photometric images, which reduces illumination-induced differences in photometric appearance and emphasizes edges (Supplementary Sec. \ref{hpf}). This two-step procedure effectively removes artifacts in the 3D height maps that would otherwise result from camera-dependent photometric appearances.

\subsection{Calibration of camera pose, distortion, and intensity variation} \label{calibration}
The first step in the 3D estimation pipeline was to calibrate the cameras' geometric and photometric properties. Specifically, the geometric properties include their 6D pose (3D position + 3D orientation) and second-order radial distortions (e.g., pincushion or barrel distortions). The photometric properties include the pixel intensity variations both within individual cameras and across different cameras. These may arise due to vignetting, uneven illumination, pixel response variation, or angle-dependent scattering of the sample.
To estimate the calibration parameters, we imaged a flat, epi-illuminated, homogeneously-patterned calibration target with the MCAM and registered the resulting 54 images, enforcing both geometric and photometric consistency in the overlapped regions.

The calibration procedure follows the optimization procedure outlined in Fig. \ref{fig:methods}a, excluding the height map-related orthorectification portion. In particular, let $\mathbf{x_0}$ and $\mathbf{y_0}$ be two vectors representing the ideal 2D spatial coordinates of the camera pixels -- that is, a 2D rectangular grid of equally-spaced points (e.g., 1536$\times$4096). Next, let $D_\theta\{\cdot,\cdot\}$ be an image deformation operation that maps from the ideal camera coordinates to a common global coordinate space (the object plane), parameterized by the camera parameters, $\theta$. See Supplementary Sec. 1 of Ref. \cite{zhou2021mesoscopic} for specific implementation details of $D_\theta$. Let $\theta_i$ be the camera parameters for the $i^\mathit{th}$ camera, so that 
\begin{equation}
    \mathbf{x_i}, \mathbf{y_i} = D_{\theta_i}\{\mathbf{x_0}, \mathbf{y_0}\}
\end{equation}
represents the (de)warped coordinates of the $i^\mathit{th}$ camera in a common object plane. 

Let $\mathbf{I}_{i,0}$ be a vector of the same length as $x_0$ and $y_0$, indicating the measured photometric intensity at every pixel coordinate for the $i^\mathit{th}$ camera. Although the debayered images have 3 color channels, here, for simplicity, we assume a single-channel image. Further, let $C_{\phi, \mathbf{x}_0, \mathbf{y}_0}\{\cdot\}$ be a photometric correction operation, parameterized by $\phi$, so that
\begin{equation}
    \mathbf{I_i} = C_{\phi_i, \mathbf{x_0}, \mathbf{y_0}}\{\mathbf{I_{i,0}}\}
\end{equation}
represents the photometrically-adjusted intensity values for the 
$i^\mathit{th}$ camera. The dependence on $\mathbf{x_0}$ and $\mathbf{y_0}$ indicates that the photometric correction is spatially-varying. Specifically, we used a second-order polynomial correction,
\begin{equation}
\begin{split}
   \mathbf{I_i} = C_{\phi_i, \mathbf{x_0}, \mathbf{y_0}}\{\mathbf{I_{i,0}}\} = 
   (&a_{i,0} + a_{i,1}\mathbf{x_0} + a_{i,2}\mathbf{y_0} + a_{i,3}\mathbf{x_0\odot x_0} + \\
   &a_{i,4}\mathbf{y_0\odot y_0} + a_{i,5}\mathbf{x_0\odot y_0})\odot\mathbf{I_{i,0}},
\end{split}
\end{equation}
where $\odot$ represents element-wise multiplication and $\phi_i=\{a_{i,0},a_{i,1},a_{i,2},a_{i,3},a_{i,4},a_{i,5}\}$. In sum, assuming $\theta_i$ and $\phi_i$ are optimized, then $\{\mathbf{x_i}, \mathbf{y_i}, \mathbf{I_{i,0}} \}$ represents the corrected $i^\mathit{th}$ camera data, accounting for distortion and photometric variation.

Next, let $\{\mathbf{x}, \mathbf{y}, \mathbf{I} \}$ be three vectors representing the flattened concatenation of $\{\mathbf{x_i}, \mathbf{y_i}, \mathbf{I_{i,0}} \}$ for all $i$. We then initialize a blank matrix, $\mathbf{R}[\cdot,\cdot]$, representing the stitched reconstruction, into which we backproject the collection of points,
\begin{equation}\label{recon_eq}
    \mathbf{R}[\mathbf{x}, \mathbf{y}] \leftarrow \mathbf{I},
\end{equation}
with interpolation, as $\mathbf{x}$ and $\mathbf{y}$ are continuously valued. When specific coordinates are visited more than once, the values are averaged. The result of Eq. \ref{recon_eq} is an estimate of the stitched composite for a given set of $\{\theta_i,\phi_i\}_{i=1}^{54}$. To update these parameters, we form a forward prediction from $\mathbf{R}[\cdot,\cdot]$ by reprojecting back into the camera spaces, as follows:
\begin{equation}
    \mathbf{I}_\mathit{pred}=\mathbf{R}[\mathbf{x}, \mathbf{y}].
\end{equation}
$\mathbf{I}_\mathit{pred}$ should match $\mathbf{I}$ when the camera images are well-registered and the corrected photometric intensities match in overlapped regions. Thus, we minimize the error metric,
\begin{equation}
    \mathit{MSE}=\|\mathbf{I}_\mathit{pred} - \mathbf{I}\|^2,
\end{equation}
with respect to $\{\theta_i,\phi_i\}_{i=1}^{54}$ via gradient descent. Since the image target is homogeneous, we also include a regularization term,
\begin{equation}
    \sum_i \mathit{stdev}(\mathbf{I_i}),
\end{equation}
which enforces a homogeneous reconstruction. Here, the standard deviation ($\mathit{stdev}$) is taken across all the pixels in one image.

Finally, we apply the calibrated parameters, $\{\theta_i,\phi_i\}_{i=1}^{54}$, to each frame of the videos of the freely-moving organisms. To homogenize the background in the case of zebrafish, which uses transmission illumination instead of the epi-illuminated calibration target, we apply a second calibration step that only optimizes the photometric correction parameters, $\{\phi_i\}_{i=1}^{54}$, using the maximum projection of the video across time, which eliminates all moving objects.

\subsection{Organism tracking and pose determination} \label{tracking}
To track the fruit flies, zebrafish larvae, and harvester ants, we first thresholded the photometric composites to segment each organism and compute each of their centroids across all video frames. We then employed a simple particle-tracking algorithm, matching the organisms by finding the closest centroid in the subsequent video frame. In the case of clashing match proposals, we assigned matches that minimized the sum of the total absolute lateral displacements. To track the ants' 6 femur-tibia joints, we incorporated the observation that the joint heights are local maxima in the 3D height maps for segmentation, and employed a similar particle-tracking algorithm.

To determine the orientation of the organisms, we performed principal component analysis (PCA) on the thresholded pixel coordinates and took the first principal component (PC) as the organism's orientation. In the case of zebrafish, we used the height map coordinates to perform PCA in 3D, thereby allowing us to compute the elevation angles in Fig. \ref{fig:zebrafish}. We resolved the sign ambiguity of the PC either by enforcing the dot products of PCs of the tracked organism in consecutive frames to be positive, or by computing the relative displacement between the unweighted centroid and the intensity-weighted centroid and forcing the PC to point in the same direction.

The fish eye vergence angles were estimated by thresholding the green channel of the photometric intensity images to identify the eyes. The orientations of the eyes were estimated using the \texttt{regionprops} command in MATLAB, which finds the angle of the major axis of the ellipse with the equivalent second moments. The vergence angle is then computed as the angle between the two eyes.

\subsection{Biological samples and data acquisition}
Zebrafish stocks were bred and maintained following IACUC guidelines and as previously described \cite{westerfield2000zebrafish}. Zebrafish were stored at 28°C with daily feeding and water changes, and cycled through 14 hours of light and 10 hours of darkness per day. Free swimming fish were imaged at larval stages between 5 dpf and 20 dpf. Specifically, zebrafish larvae were transferred from culture chambers using a transfer pipette to a clear plastic imaging arena (with lateral inner dimensions 97 mm × 130 mm), which was filled with system water a few mm deep. The arena was then placed on the sample stage of the MCAM system. The z position of the stage was adjusted such that the zebrafish larvae were all within the DOF of the lenses. The system was left undisturbed with the LED illumination panels turned on for at least 5 minutes to allow the zebrafish to acclimate, after which multiple MCAM videos were acquired using a custom Python script. After video acquisition, the arena was removed and replaced with a flat patterned calibration target. We focused the target with the z stage using a Laplacian-based sharpness metric and captured a single frame (all 54 cameras), which would serve to calibrate the camera poses and distortions for all videos captured during that imaging session.

The wild-type red harvester ants and fruit flies (available from various vendors on Amazon) were maintained at room temperature. When ready for imaging, we positioned and focused a flat patterned calibration target, which serves two purposes: 1) for camera calibration, just as for the zebrafish videos described in the previous paragraph, and 2) to serve as a flat substrate for the ants and fruit flies to walk upon. The patterned target, although not required, serves as a global reference in the 3D height maps. Alternatively, the substrate could be monochrome/featureless or transparent (e.g., a glass sheet), as was the case for the zebrafish imaging configuration, in which case the 3D height map would assign an arbitrary height value to the background without affecting the 3D accuracy of the organisms themselves. 

The ants or fruit flies were inserted into a Falcon tube and released onto the center of the flat substrate, after which we immediately ran the same custom Python script to acquire MCAM videos. If necessary, the insects were re-collected in the tubes and re-released into the arena for repeated imaging. After video acquisition, we acquired a single frame of the calibration target alone, just as we did after zebrafish video acquisition.

\newpage
\bibliography{sample}

\section*{Acknowledgements}
We would like to thank Kristin Branson, Srinivas Turaga, Timothy Dunn, Archan Chakraborty, and Maximilian Hoffmann for their helpful feedback on the manuscript. Research reported in this publication was supported by the Office of Research Infrastructure Programs (ORIP), Office Of The Director, National Institutes Of Health of the National Institutes Of Health and the National Institute Of Environmental Health Sciences (NIEHS) of the National Institutes of Health under Award Number R44OD024879, the National Cancer Institute (NCI) of the National Institutes of Health under Award Number R44CA250877, the National Institute of Biomedical Imaging and Bioengineering (NIBIB) of the National Institutes of Health under Award Number R43EB030979, the National Science Foundation under Award Number 2036439, and the Duke Coulter Translational Partnership Award.

\section*{Author contributions}
KCZ and RH conceived the idea and initiated the research. KCZ developed the algorithms and theory, with the help of CLC, JP, PCK, and RH. KCZ wrote the code for and performed 3D video reconstruction and stitching, animal tracking, and data analysis. MH, JD, PR, VS, CBC, MZ, and RH developed the MCAM hardware and acquisition software. KCZ acquired and analyzed the biological data, with the help of JPB, JB, AB, GH, and RH. JD and KCZ created the supplementary videos. KCZ wrote the manuscript and created the figures, with input from all authors. RH and MB supervised the research.

\section*{Disclosures}
RH and MH are cofounders of Ramona Optics, Inc., which is commercializing multi-camera array microscopes. MH, JP, JD, PR, VS, CBC, MZ, JPB, and GH are or were employed by Ramona Optics, Inc. during the course of this research. KCZ is a consultant for Ramona Optics, Inc.

\section*{Data availability}
Data will be available at https://doi.org/10.7924/r4db86b1q.

\section*{Code availability}
Code will be available at https://github.com/kevinczhou/3D-RAPID.

\newpage

\renewcommand{\thesection}{S\arabic{section}}
\renewcommand{\thetable}{S\arabic{table}}
\renewcommand{\thefigure}{S\arabic{figure}}
\renewcommand\theequation{S\arabic{equation}}
\setcounter{figure}{0}
\setcounter{table}{0}
\setcounter{section}{0}
\setcounter{equation}{0}

\section*{Supplementary information}

\section{System characterization: lateral resolution, axial precision and accuracy, and depth of field} \label{supp_characterization}

\begin{figure}[h]
    \centering
    \centerline{\includegraphics[width=1.2\textwidth]{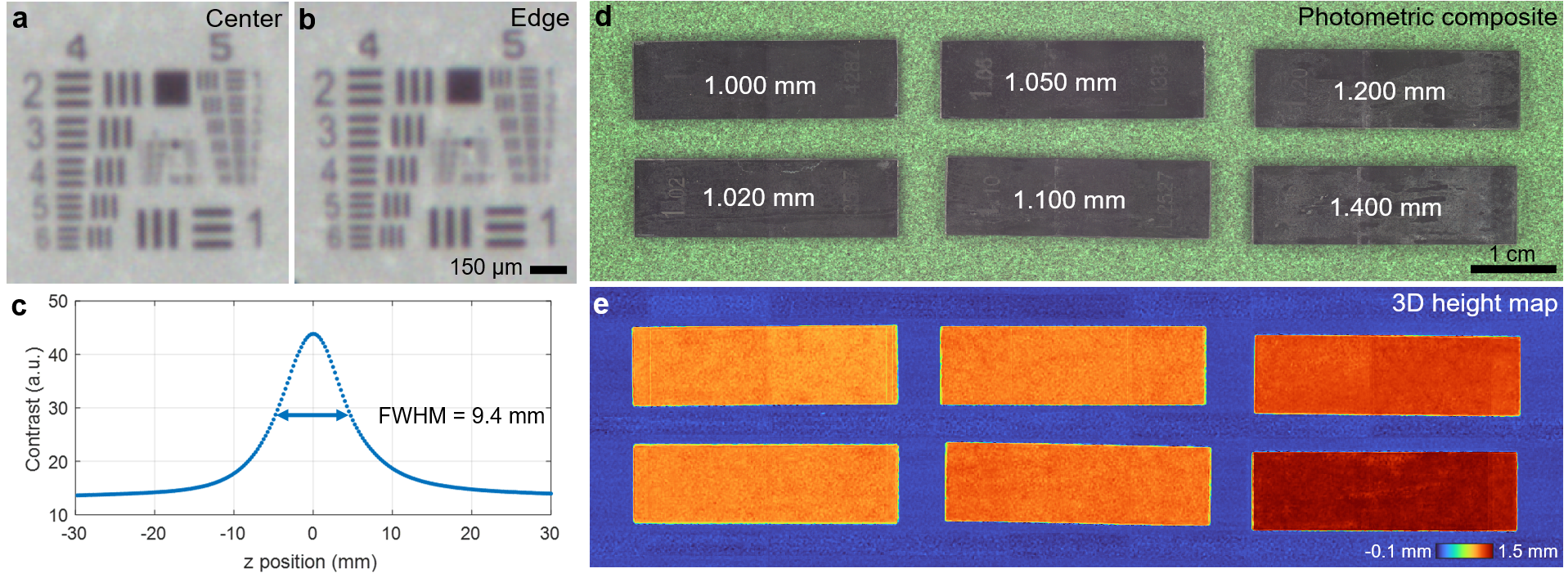}}
    \caption{System characterization experiments. \textbf{a}, \textbf{b}, USAF resolution test chart image near the center and edge of the FOV of one camera without downsampling. \textbf{c}, Image contrast of a patterned target as a function of axial position. \textbf{d}, Stitched photometric composite of 6 precisely-machined gauge blocks placed on a green patterned target (captured with the 60-fps configuration), with their nominal thicknesses denoted. \textbf{e}, The reconstructed 3D height map of the gauge blocks. Accuracy and precision are quantified in Table \ref{table:gauge_analysis}.}
    \label{fig:characterization}
\end{figure}

We performed several experiments to characterize the performance of our computational 3D imaging system, starting with imaging of a USAF resolution target near the center and edge of the field of view (FOV) of a single camera (Fig. \ref{fig:characterization}a,b). Our system can resolve group 5 elements 2-3, corresponding to a bar width of 12-13 \textmu m or a full-pitch lateral resolution of $\sim$25 \textmu m. We then characterized the depth of field (DOF) by axially translating the same flat patterned target used in Figs. \ref{fig:flies} and \ref{fig:ants}, using a motorized stage (Zaber) in increments of 0.25 mm. This defines the axial FOV of our 3D reconstructions. For each axial position, we computed a contrast metric based on the mean image gradient magnitude (Fig. \ref{fig:characterization}c). The full width at half maximum (FWHM) of this curve is 9.434 mm, which is similar to value obtained by fitting the curve to the intensity of a Gaussian beam,
\begin{equation}
    I(z) = \frac{I_0}{1+\frac{(z-z_0)^2}{z_R^2}}  + I_b,
\end{equation}
where $I_0$ and $I_b$ are the arbitrary amplitude and offset, $z_0$ is the focal position, and $2z_R$ is the DOF, corresponding to when the lateral resolution degrades by $\sqrt{2}$. Least-squares fitting yields $2z_R=9.402$ mm. In practice, the DOF may be smaller if the neighboring cameras are not focused to the same plane, such that the focus regions are offset.

Finally, we characterized the accuracy and precision of our 3D height maps by imaging 6 gauge blocks (Mitutoyo), precisely machined and characterized to be within 0.3 \textmu m of their nominal values: 1.000, 1.020, 1.050, 1.100, 1.200, and 1.400 mm (Fig. \ref{fig:characterization}d,e). We computed the accuracy as the absolute error between the estimated and ground truth heights, aggregated across all pixels within each gauge block, and the precision as the standard deviation of the height estimates across each gauge block, which are summarized in Table \ref{table:gauge_analysis} for all three configurations in Table \ref{table:configurations}. Since there is an arbitrary global height offset, we chose the one that minimizes the MSE between the estimated and ground truth heights \cite{zhou2021mesoscopic}.

\begin{table}
\centering
\begin{tabular}{ c | c c | c c | c c}
Ground truth & \multicolumn{2}{c|}{1$\times$ downsamp} & \multicolumn{2}{c|}{2$\times$ downsamp} & \multicolumn{2}{c}{4$\times$ downsamp}\\
 height & Acc. & Prec. & Acc. & Prec. & Acc. & Prec.\\ 
 \hline
 0    & 44.3 & 19.3 & 25.3 & 17.2 & 60.0 & 55.9\\  
 1000 &  8.9 & 17.5 & 12.0 & 32.2 & 50.6 & 69.4\\  
 1020 &  4.1 & 11.2 & 18.1 & 24.3 & 51.6 & 72.7\\  
 1050 &  3.2 & 18.5 &  4.3 & 25.7 & 14.8 & 63.8\\  
 1100 &  7.9 & 17.7 &  7.8 & 28.6 & 12.7 & 68.7\\  
 1200 &  5.2 & 24.1 &  0.4 & 33.0 & 20.3 & 88.9\\  
 1400 & 55.0 &  8.7 &  1.0 & 27.7 & 49.4 & 100.4\\  
 \hline
 mean & 18.4 & 16.7 &  9.8 & 26.9 & 37.1 & 74.3
\end{tabular}
\caption{Accuracy (absolute error from ground truth) and precision (standard deviation) of the height estimation of the 6 gauge blocks (and background) in Fig. \ref{fig:characterization}a,b for all three downsampling configurations. All values are in \textmu m.}
\label{table:gauge_analysis}
\end{table}

\section{Generalization experiments}\label{monocular}
Here, we show that the multiocular stereo CNN trained on a subset of frames can generalize well to unseen frames. As validation we compare this generalization performance to that of a monocular stereo CNN (i.e., one that only takes in a single image as the input). To make these comparisons, we picked two independent subsets of the video frames. In Set 1, we took about 15 frames equally spaced temporarily across the video. In Set 2, we took another 15 equally spaced frames at half a period offset with respect to Set 1. For example, if the video was 601 frames, then Set 1 would consist of frames 1, 41, 81, ... 561, 601 and Set 2 would consist of frames 20, 60, 100, ...540, 580. We then trained two independent multiocular CNNs, one on Set 1, the other on Set 2, and compared the 3D height map predictions on both sets. The idea is that in the absence of ground truths, the physics-supervised CNN predictions on training set examples could serve as pseudo-truths. For comparison, we also trained a monocular CNN on Set 1 and compared predictions on Set 1 and Set 2. 

\begin{figure}[ht]
    \centering
    \centerline{\includegraphics[width=\textwidth]{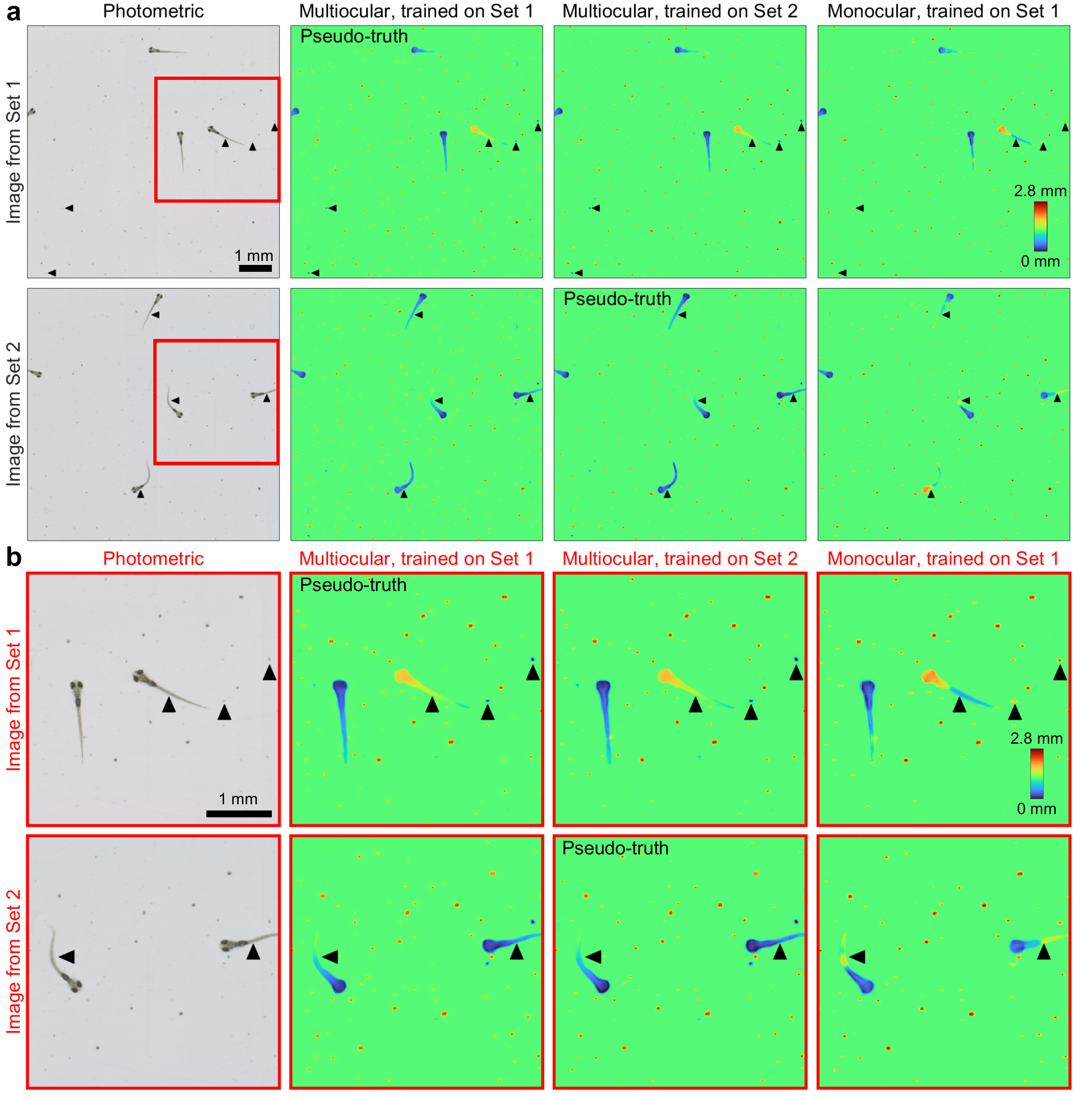}}
    \caption{Generalization performance of multiocular and monocular CNNs trained on frames from a video of freely swimming zebrafish. \textbf{a}, First row shows an example from Set 1 and 3D height predictions of three different CNNs -- two multiocular CNNs, trained on Set 1 and Set 2, and one monocular CNN trained on Set 1. Second row shows predictions on Set 2. \textbf{b}, Zoom-in of the red boxes in \textbf{a}. Arrowheads point out features for which the multiocular CNN generalized well, but not the monocular CNN, as evaluated by comparing the predictions the respective pseudo-truth.}
    \label{fig:generalization_fish}
\end{figure}

\begin{figure}[ht]
    \centering
    \centerline{\includegraphics[width=1.2\textwidth]{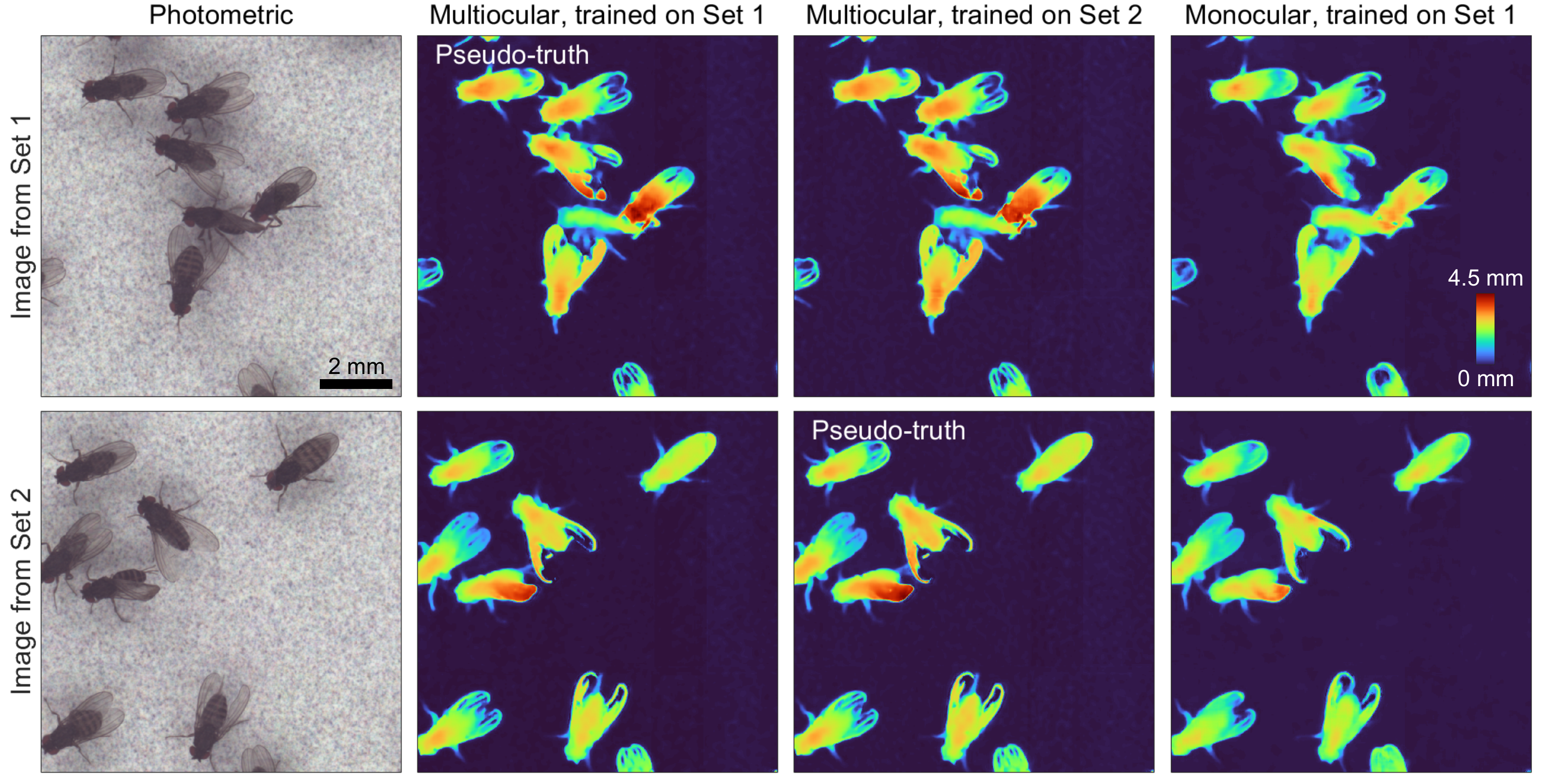}}
    \caption{Generalization performance of multiocular and monocular CNNs trained on frames from a video of fruit flies. First row shows an example from Set 1 and 3D height predictions of three different CNNs -- two multiocular CNNs, trained on Set 1 and Set 2, and one monocular CNN trained on Set 1. Second row shows predictions on Set 2.}
    \label{fig:generalization_flies}
\end{figure}

Figs. \ref{fig:generalization_fish} and \ref{fig:generalization_flies} show the comparisons among these three CNNs for both zebrafish and fruit flies. In both organisms, the multiocular CNNs generalize well to unseen video frames, 
based on comparisons
between images from the CNN trained on Set 1 and the one trained on Set 2. However, for zebrafish, the monocular CNN (trained on Set 1) generalizes poorly (to Set 2). This is evidenced by erroneous heights of several zebrafish's heads or tails, as it is difficult to determine the heights of the fish based on appearance alone -- magnification-based cues are confounded by natural size variation. Similarly, the monocular CNN incorrectly estimates the heights of the sunken food particles. This is likely due to the fact that the vast majority of food particles are floating, and since the food particles have no discernible height indicators, the monocular CNN simply uniformly assigns the floating height to all particles. While the monocular CNN performs better for the fruit flies than for zebrafish, it still makes a few errors, e.g., when one fly is climbing on top of another. Such fly behavior was rare in our captured video, so the monocular CNN had fewer training examples to learn the semantic cues to accurately predict the elevated height, whereas the multiocular CNN was able to predict the elevated height from the parallax cues.

\section{Implementation details on patch-based training with multi-ocular stereo inputs} \label{patch_detailed}
Here, we expand upon the explanation of our patch-based CNN training procedure given in Sec. \ref{patch} and Fig. \ref{fig:methods}c. 

\subsection{Determining the observing cameras and the coordinates}
We start with the camera pose calibration based on a flat patterned target (Methods \ref{calibration}) to generate a ``visitation log'', $V$. $V$ is an $n_\mathit{row}$ $\times$ $n_\mathit{column}$ $\times$ 54 $\times$ 2 tensor look-up table specifying which of the 54 cameras view a certain spatial position in the reconstruction coordinate system as well as the respective (row,column) pixel coordinates in the camera coordinate system that map to that position. The formation process of $V$ is somewhat similar to the backprojection step of the reconstruction (Fig. \ref{fig:methods}a), but instead of backprojecting the RGBH values, we backproject the (row, column) coordinates. 
This visitation log facilitates rapid retrieval of the relevant cameras for each randomly sampled position. Note that since we want to avoid rolling shutter artifacts that may occur where the bottom of one camera overlaps with the top of the camera below (Methods \ref{rolling_shutter} and Supplementary Sec. \ref{rolling_shutter_supp}), we only consider horizontal overlap.

\subsection{Selecting random patches}

Given this visitation log, we select $n_\mathit{batch}$ random 2D coordinates in the reconstruction frame of reference for each CNN training iteration. For each of these random coordinates, we retrieve the relevant cameras and their corresponding camera-centric coordinates. For each camera image, we then crop out a square patch of width $w_\textit{patch}$ centered at the sampled coordinates. If these coordinates are within $w_\textit{patch}/2$ of a camera image edge, they are shifted so that the patch remains within the image. 

For each image patch, we also extract patches from the left and right cameras and stack them along the channel dimension of the CNN input, which the CNN can exploit for 3D estimation (Fig. \ref{fig:methods}c). To do this in a manner consistent with both training on patches and inference on full-sized images, we homographically transformed the left/right neighboring images into the frame of reference of the central camera in question, as if the sample were flat (more precisely, coincident with the pre-calibration reference plane; Sec. \ref{stitching}, Methods \ref{calibration}). If the sample were completely flat, then the transformed neighboring images would theoretically be identical to the image captured by the camera in question where their viewpoints overlap. However, if the sample exhibits height variation, the transformed neighboring images would exhibit parallax shifts in proportion to the height variation. When there is no left or right camera (i.e, the first or last column of cameras), we input blank images (all zeros). Similarly, when either the left or right patch overlaps with the edge of its respective camera, we assign zeros to the missing regions. Note that in this scenario, we cannot shift the left/right patch away from the edge, as we could above, because the left/right patch must remain coaligned with the main (central) patch so that we maintain full convolutionality for the inference step (Supplementary Sec. \ref{inference}). Furthermore, we do not want to exclude training cases where the central patch is close to the edge of the camera, as these cases appear when applied to full-size camera images during the inference step.

We note that the number of cameras observing a particular point can range from 1 - 3, since we only consider horizontal overlap. When only one camera views a particular point (the left and right edges of the reconstruction) during training, we reject the resulting patch as there's nothing to register. To account for the fact that the number of patches may vary for each batch element, we use tensorflow's \cite{abadi2016tensorflow} \texttt{tf.RaggedTensor} construct, which allows some dimensions of a tensor to have slices with different lengths. In our experiments, we used $n_\mathit{batch}=$1, 2, and 8 for the $1\times$, $2\times$, and $4\times$ downsampling cases.

\subsection{CNN architecture}
The input to the CNN has nine channels, corresponding to three stacked RGB inputs -- the camera image whose height we wish to predict, followed by the left and right camera views (Fig. \ref{fig:methods}c). The output of the CNN is a single-channel height map, obtained by summing across the channel dimension of the final convolutional layer.

The encoder-decoder CNN architectures were based on one basic building block, consisting of the following operations in sequence:
\begin{enumerate}
\setlength{\itemsep}{0pt}
    \item $3\times3$ convolution, $k$ filters, stride=1, padding=`same',
    \item Batch normalization,
    \item Leaky ReLU,
    \item $1\times1$ convolution, $k$ filters, stride=1, padding=`valid',
    \item Batch normalization,
    \item Leaky ReLU (unless final block of the CNN),
\end{enumerate}
where $k$ is a free hyperparameter, specifying the number of filters in the convolution layers. In the case of an upsample block, a $2\times$ nearest-neighbor upsampling procedure is applied \textit{before} the block. In the case of a downsample block, a 2$\times$2 max pooling operation is applied \textit{after} the block. 

The full, symmetric encoder-decoder CNN architecture is described by a list of positive integers, each of which specifies the $k$ for an upsample/downsample block pair. For example, [8, 16, 32] indicates three downsample blocks with $k$ = 8, 16, and 32 filters, followed by three upsample blocks with $k$ = 32, 16, and 8 filters. In our experiments, we set $k$ = 32 for all upsample/downsample blocks, but varied the number of blocks between 3 and 6 (i.e., [32, 32, 32] and [32, 32, 32, 32, 32, 32]), depending on the sensor downsampling.

\subsection{Data-dependent loss function}
The data-dependent loss function is computed based on the model depicted in Fig. \ref{fig:methods}a, where 2-3 image patches are used instead of 54 full-size images. Specifically, the 4-channel (RGBH) image patches are backprojected onto a blank ``canvas'' according to the camera poses and height map-derived orthorectification fields (Eq. \ref{orthorectification}). The same coordinates are then used to reproject back to camera-centric coordinates to obtain the forward predictions. The data-dependent loss function is thus the MSE between forward predictions and the original RGBH patches.

\subsection{Normalized high-pass filtering}\label{hpf}
For terrestrial samples, which were illuminated in reflection, we found that registering the RGB images sometimes led to artifacts due to camera-dependent photometric appearance. This can be caused by illumination variation across the FOV due to off-axis LED panel geometry and anisotropic, non-Lambertian reflections, causing different amounts of light entering each camera. To combat these effects, we used normalized high-pass filtered versions of the images,
\begin{equation}\label{eq:hpf}
    \widetilde{I}_\sigma(x,y)=
    \frac{I(x,y)\circledast\exp\left(-\frac{x^2+y^2}{4\sigma^2}\right)}
    {I(x,y)\circledast\exp\left(-\frac{x^2+y^2}{2\sigma^2}\right)},
\end{equation}
where $\circledast$ denotes 2D convolution. Thus, Eq. \ref{eq:hpf} is the ratio of two Gaussian-blurred versions of $I(x,y)$, the grayscale-converted RGB image, with widths $\sigma$ and $\sqrt{2}\sigma$. Like high-pass filtering, applying Eq. \ref{eq:hpf} to the images highlights edges and attenuates DC and low-frequency features. The motivation for taking a ratio rather than subtracting (i.e., difference of Gaussians) is so that the spatial fluctuations are normalized and therefore illumination-variation-independent, thereby facilitating registration. To capture different scales, we used three values of $\sigma$ for the three image channels ($\sigma=1, 2, 4$). 

\subsection{Regularization of the height maps}\label{height_map_regularization}
In addition to the CNN reparameterization (i.e., DIP) of the height maps as a regularizer \cite{ulyanov2018deep,zhou2020diffraction,zhou2021mesoscopic}, we also incorporated two additive regularization terms to the overall loss function: height map consistency regularization and support regularization. The height map consistency regularization enforces agreement in height values in overlapped regions of camera images and simply comes from the fourth channel of the RGBH images, whose contribution can be scaled by a hyperparameter, $\lambda_\mathit{height}$. We observed smoothing effects with increasing $\lambda_\mathit{height}$. The object support regularization relies on a segmentation mask of the background pixels, whose height values we enforce to be a particular constant (e.g., 0) via an L2 loss. In other words,
\begin{equation}\label{eq:support}
    \mathit{loss}_\mathit{support} = \lambda_\mathit{support}\sum_\mathit{x,y} \mathit{mask}_\mathit{background}(x,y) (h(x,y)-h_0)^2,
\end{equation}
where $\mathit{mask}_\mathit{background}(x,y)$ is the segmentation mask, $h(x,y)$ is the height map output of the CNN,  $h_0$ is the known background height value, and $\lambda_\mathit{height}$ is the regularization coefficient. In this paper, we used a simple intensity-based threshold on the green channel of the photometric images, as our backgrounds are relatively homogeneous, although other segmentation strategies may be used.

\subsection{Additional training details}
We optimized the loss function, consisting of the aforementioned data-dependent and regularization terms, using the Adam Optimizer \cite{kingma2014adam}. Depending on the downsampling configuration, we used a different patch size and number of patches per iteration: one 1024$\times$1024 patch (no downsampling), two 768$\times$768 patches ($2\times$ downsampling), and eight 384$\times$384 patches ($4\times$ downsampling). These patches were randomly selected from a subset of the recorded video frames -- for the $2\times$ and $4\times$ downsampling configurations, we selected from 15-16 frames evenly distributed frames, while for the no downsampling configuration, we used 8 frames (due to memory constraints).  

For the reflection-illuminated terrestrial samples, we performed a two-step training procedure, where we first optimized with RGB images using $\lambda_\mathit{height}=500$ (Supplementary Sec. \ref{height_map_regularization}) to scale the height channel (with units of mm) and $\lambda_\mathit{support}=0$ (Eq. \ref{eq:support}) for 30k iterations. Thereafter, we ran 70k iterations with the normalized high-pass filtering (Supplementary Sec. \ref{hpf}) and $\lambda_\mathit{height}=50$ and $\lambda_\mathit{support}=100$. For aquatic samples, high-pass filtering was not necessary because they were illuminated in transmission. Thus, we used a one-step training procedure with 70k iterations with $\lambda_\mathit{height}=50$ and $\lambda_\mathit{support}=100$.

\subsection{Inference step - generating the full-size RGBH videos} \label{inference}
\begin{figure}[ht]
    \centering
    \centerline{\includegraphics[width=1.2\textwidth]{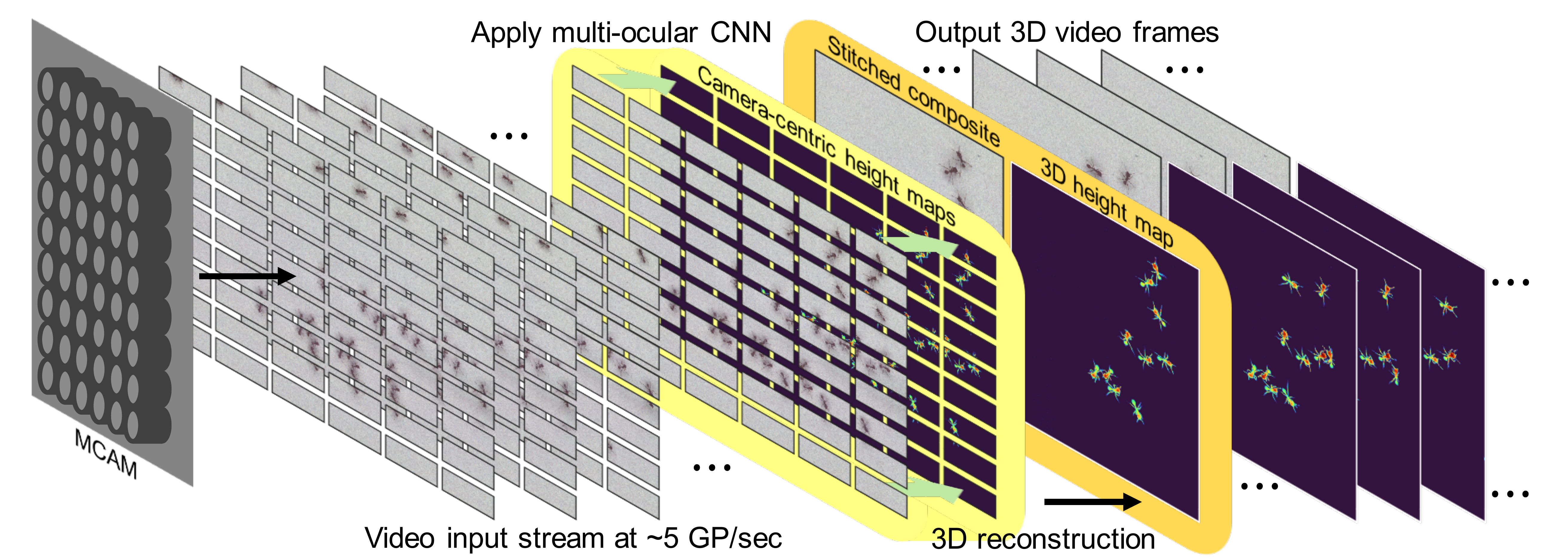}}
    \caption{Inference step post patch-based training (Fig. \ref{fig:methods}c) that generates the stitched composites and coregistered 3D height map on potentially unseen video frames.}
    \label{fig:cnn_inference}
\end{figure}

Once the CNN is trained to map from multi-ocular stereo inputs to a 3D height map using the patch-based procedure, we can apply the CNN to sequences of full-sized MCAM video streams that includes unseen frames (Fig. \ref{fig:cnn_inference}). Essentially, this refers to the backprojection step in Fig. \ref{fig:methods}a. Since iterative optimization is no longer necessary after the CNN is fully trained, generating new 3D video frames can be done quickly. For example, one application might involve a human observer selecting a particular region of interest within the large FOV, whose 3D height map the computer would then generate in real time.

\section{Reducing the impact of the per-camera rolling shutter}\label{rolling_shutter_supp}
Each sensor exhibits a rolling shutter, whereby the pixels begin integrating sequentially every $\delta t = (230\ \mathrm{MHz})^{-1} = 4.35$ ns and are read out in a raster scan pattern row by row from the top left to bottom right (with the longer sensor dimension as the horizontal dimension). Although the rolling shutters are synchronized to within 10 \textmu s across cameras, there is still significant asynchrony in overlapped regions of neighboring camera FOVs, thus thwarting accurate 3D estimation. Here, we consider asynchrony in 1) vertically overlapped FOVs and 2) horizontally overlapped FOVs. The former asynchrony is much more serious, as the bottom row of the upper sensor is not reached until after $\delta t \times l_\mathit{row} \times l_\mathit{col}$, where $l_\mathit{row}$ and $l_\mathit{col}$ are the number of pixels per row and column, respectively. Using the full sensor without downsampling ($l_\mathit{row}=4208$, $l_\mathit{row}=3120$), the time delay between the last row of the upper sensor and the first row of the lower sensor is $\sim$57 ms. In practice, the delay is even larger due to horizontal and vertical blanking (dead time between row and column reads). To circumvent this problem, we thus reduced the number of rows approximately in half (3120 to 1536) to ensure the smallest overlap between vertically adjacent cameras that still allowed for a contiguous composite FOV. This also has the added benefit of increasing the sensor frame rate.

Asychrony in horizontally overlapped FOVs is less serious, but still an important consideration. Using the full sensor without downsampling, the time delay between corresponding rows of perfectly aligned camera FOVs is only $\delta t \times l_\mathit{row}$, or approximately 20 \textmu s, which is negligible. In practice, however, there is a vertical offset due to slight camera misalignments, so that the time delay is $\delta t \times l_\mathit{row} \times l_\mathit{misalign}$. Based on stitching a flat target, we determined that the worst-case vertical misalignment was $l_\mathit{misalign}$ = 100 rows, leading to a 2-ms delay between when the corresponding pixels in horizontally neighboring cameras begin to expose. To ensure significant temporal overlap (at least 90\%) in the exposure periods, we thus exposed for 2 ms$/(1-0.9) = $ 20 ms.

For $2\times$ and $4\times$ downsampling, the asynchrony is less dramatic because the numbers of rows and columns are reduced. Going through similar calculations, we determined that exposing for 5 ms and 2.5 ms for $2\times$ and $4\times$ downsampling, respectively, leads to $>$90\% temporal overlap in the worst-case vertical camera misalignment cases. Note that these values don't quite scale proportionally between the $2\times$ and $4\times$ cases due to horizontal blanking periods not decreasing proportionally.

\section{Impact of hardware design on height accuracy}
\label{hardware_theory}

\begin{figure}[ht]
    \centering
    \includegraphics{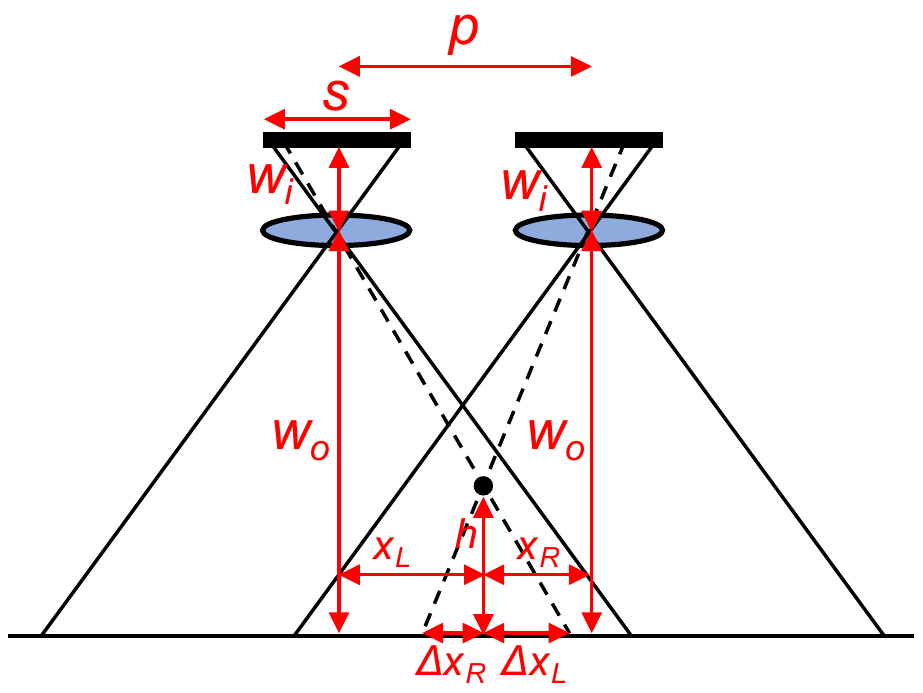}
    \caption{Two identical cameras with effective focal length $f$ observing a common sample point with height $h$ from the focal plane. The magnification is $M=w_i/w_o$.}
    \label{fig:height_accuracy}
\end{figure}

Here, we explore how hardware design choices impact the accuracy of 3D height estimation. We will ignore errors stemming from camera distortion, aberrations, and misalignment and assume ideal paraxial imaging performance. Further, for simplicity, we assume two adjacent cameras spaced by $p$ center to center with a common effective focal length, $f$, a working distance (i.e., the distance between the sample plane and the lens principal plane) of $w_o$, and a sensor-to-lens distance of $w_i$ (Fig. \ref{fig:height_accuracy}). These latter three parameters satisfy the lens equation, 
\begin{equation}
    \frac{1}{w_o}+\frac{1}{w_i} = \frac{1}{f}.
\end{equation}
The magnification is thus $M=w_i/w_o$.

Further, consider a sample point with height $h$ positioned $x_L$ from the optical axis of the left camera and $x_R$ from that of the right camera. Due to nontelecentric optics, the apparent object-side position of this sample point is parallax-shifted $\Delta x_L$ in the left camera and $\Delta x_R$ in the right camera. These shifts are related to the height via Eq. \ref{orthorectification},
\begin{equation}
    \Delta x_L = \frac{h x_L M}{f(M+1) - hM}, 
    \ \
    \Delta x_R = \frac{h x_R M}{f(M+1) - hM}.
\end{equation}
We are interested in the total parallax shift between both cameras, given by
\begin{equation}\label{delta_x}
    \Delta x = \Delta x_L + \Delta x_R = \frac{h p M}{f(M+1) - hM},
\end{equation}
which does not depend on the lateral position of the sample point, as $x_L+x_R=p$. How well we can estimate $\Delta x$ depends on how accurately we can match and register the sample point in both camera images, which in turn depends on the lateral resolution of the imaging system. We consider two limits: the diffraction-limited regime and the pixel-size-limited regime. Let $\delta x_\mathit{pixel}$ be the camera pixel size, so that $\delta x_\mathit{pixel}/M$ is the object-side pixel size. Further, let $\delta x_\mathit{diff}$ be the camera-side diffraction-limited spot size, so that $\delta x_\mathit{diff}/M$ is the object-side diffraction-limited spot size:
\begin{equation}
    \delta x_\mathit{diff} \propto \frac{\lambda}{\mathit{NA}} \approx \frac{2\lambda w_i}{w} = \frac{2\lambda f (M+1)}{w},
\end{equation}
where $w$ is the lens aperture diameter and $\lambda$ is the wavelength. Assuming that we can match corresponding points in the two camera images with an uncertainty proportional to the lateral resolution, then the corresponding height error can be estimated by setting $\Delta x$ (Eq. \ref{delta_x}) equal to the object-side lateral spot size and solving for $h$. In the pixel-resolution-limited regime ($\delta x_\mathit{pixel} \gg \delta x_\mathit{diff}$), we have that the height uncertainty is
\begin{equation}
    \delta h_\mathit{pixel} \propto \frac{f\delta x_\mathit{pixel}(M+1)}{ M(\delta x_\mathit{pixel} + p M)},
\end{equation}
meaning that downsampling the images results in a roughly proportional decrease in height uncertainty.
In the diffraction-limited regime, we have that
\begin{equation}
    \delta h_\mathit{diff} \propto \frac{2\lambda f^2 (M+1)^2}{M(2\lambda f (M+1) + p w M)}.
\end{equation}
We can see that in both cases, all else equal, decreasing $f$ and increasing $p$ and $M$ improve the height estimation accuracy. It may appear helpful to decrease $M$ to increase the amount of overlap of neighboring camera FOVs until eventually non-adjacent cameras begin to overlap, resulting in larger values of $p$. However, in both pixel-limited and diffraction-limited regimes, $1/p$ decreases more slowly than the factors that include $M$ increase as $M$ decreases (e.g., consider $p\rightarrow 2p$, $M\rightarrow M/2$). Furthermore, this analysis assumes that the object height variation is within the depth of field of the imaging systems, within which the lateral resolution remains roughly constant. Thus, while designs that increase the lateral resolution can improve height estimation accuracy, they also compromise the axial FOV.

We now consider the case where the camera FOVs are critically overlapped at 50\%, that is when $M=s/2p$, where $s$ is the sensor width. Thus, the height uncertainties in the pixel- and diffraction-limited regimes are, respectively,
\begin{equation}
    \delta h_\mathit{pixel} \propto \frac{2\delta x_\mathit{pixel} f (2p+s)}{s(2 \delta x_\mathit{pixel} + s)}
    \approx \frac{2\delta x_\mathit{pixel} f (2p+s)}{s^2},
\end{equation}
\begin{equation}
    \delta h_\mathit{diff} \propto \frac{2 \lambda f^2 (2p + s)^2}{s (2 \lambda f (2p + s) + psw)}
    \approx \frac{2 \lambda f^2 (2p + s)^2}{ps^2w}.
\end{equation}
In the ideal case of $p=s$, so that there are no gaps in between the sensors and $M=1/2$, we have
\begin{equation}
    \delta h_\mathit{pixel} \propto \frac{\delta x_\mathit{pixel} f}{p},
\end{equation}
\begin{equation}
    \delta h_\mathit{diff} \propto \frac{\lambda f^2}{pw}.
\end{equation}

\section{SNR considerations}
As with all imaging systems, SNR is an important metric for 3D-RAPID. Specifically, the better the SNR of the photometric images, the higher the image registration accuracy and by extension the 3D estimation accuracy. There are several trade offs involving SNR with our method as it relates to imaging small model organisms.
\begin{enumerate}
    \item Numerical aperture (NA): the higher the NA, the more light collected and the better the shot-noise-limited SNR. The associated improved lateral resolution also improves the 3D height estimation accuracy, because the parallax estimation accuracy would increase (Supplementary Sec. \ref{hardware_theory}). However, at the same time, the higher the NA, the shallower the depth of field, which limits the axial FOV of the 3D reconstructions. In addition, the higher the NA, the smaller the lateral FOV becomes in practice due to difficulties in correcting aberrations \cite{zheng2014fourier} and therefore the tighter the camera array packing would need to be.
    \item Behavior: while increasing the illumination power would yield higher SNR, care must be taken to avoid influencing the behavior of the model organisms. This tradeoff can be partially alleviated by using wavelengths invisible to the model organism’s visual system, however radiative heating from the illumination source can potentially still influence behavior.
    \item Speed: the higher the frame rate, the less light that is detected and therefore the lower the SNR per frame. Increasing illumination power can alleviate this tradeoff until it influences the behavior of interest.
    \item Camera type: one of the factors enabling the financial tractability of the 3D-RAPID architecture is its use of CMOS digital image sensors that are currently fabricated at large scales for the cell phone camera market. While the sensitivities of these camera sensors have improved significantly over the past decade (e.g., now with very low read noise and dark current and high quantum efficiency, due in part to the introduction of back-side illuminated CMOS sensors), their performance may still generally lag behind that of high-end scientific CMOS and EMCCD sensors. While this latter technology is currently too expensive to multiplex into an array with more than several dozen sensors, it may become feasible in the future.
\end{enumerate}

\section{Supplementary video descriptions}
\begin{enumerate}
    \item 60-fps, 36.6-MP video of freely swimming zebrafish larvae (10 dpf) feeding on mostly floating AP100 food particles. The left panel is the photometric composite and the right panel is the 3D height map. The video zooms into three feeding events (or attempts) by two different fish.
    \item 230-fps, 9.1-MP video of freely swimming zebrafish larvae (10 dpf) feeding on mostly floating AP100 food particles. The left panel is the photometric composite and the right panel is the 3D height map. The video zooms in on three independent feeding events by three different fish. The third fish can be seen swallowing the food particle.
    \item 60-fps, 36.6-MP video of freely swimming zebrafish larvae (10 dpf) feeding on mostly floating AP100 food particles. The left panel shows the full field of view with the trajectories mapped out. The panels on the right each correspond to individual fish, uniquely identified by a 2-digit number, whose position and orientation are denoted with red annotations. The righthand panels' border colors nonuniquely match those of the tracks in the lefthand panel, to assist the viewer in matching the fish to the trajectories. Righthand panels appear and disappear when the fish enters or exits the FOV. The first half of the video shows the photometric values, while the second half of the video shows the 3D height maps.
    \item 60-fps, 36.6-MP video of 20-dpf zebrafish larvae feeding on live brine shrimp. The left panel is the photometric composite and the right panel is the 3D height map. The video zooms in on two feeding events from two different fish.
    \item 230-fps, 9.1-MP video of 20-dpf zebrafish larvae feeding on live brine shrimp. The left panel is the photometric composite and the right panel is the 3D height map. The video zooms into one feeding event.
    \item 60-fps, 36.6-MP video of a large school of 5-dpf zebrafish larvae freely swimming in an open arena at high speed. The left panel is the photometric composite and the right panel is the 3D height map.
    \item 230-fps, 9.1-MP video of a large school of 5-dpf zebrafish larvae freely swimming in an open arena at high speed. The left panel is the photometric composite and the right panel is the 3D height map.
    \item 60-fps, 36.6-MP video of freely moving fruit flies. The left panel is the photometric composite and the right panel is the 3D height map.
    \item 230-fps, 9.1-MP video of freely moving fruit flies. The left panel is the photometric composite and the right panel is the 3D height map. 
    \item 60-fps, 36.6-MP video of freely moving fruit flies. The left panel shows the full field of view with the trajectories mapped out. The panels on the right each correspond to individual flies, uniquely identified by a 2-digit number, whose position is denoted by a red circle. The righthand panels' border colors nonuniquely match those of the tracks in the lefthand panel, to assist the viewer in matching the flies to the trajectories. Righthand panels appear and disappear when the fish enters or exits the FOV. The first half of the video shows the photometric values, while the second half of the video shows the 3D height maps.
    \item 60-fps, 36.6-MP video of freely moving harvester ants. The left panel is the photometric composite and the right panel is the 3D height map. 
    \item 230-fps, 9.1-MP video of freely moving harvester ants. The left panel is the photometric composite and the right panel is the 3D height map. 
\end{enumerate}

\end{document}